\documentclass[aps,pre,amsmath,10pt,twocolumn,superscriptaddress]{revtex4-1}
\usepackage[dvipdfmx]{graphicx}
\usepackage{color}
\usepackage{bm}
\usepackage{multirow}

\begin{document}
\title{Linearized machine-learning interatomic potentials for non-magnetic elemental metals: Limitation of pairwise descriptors and trend of predictive power}
\author{Akira \surname{Takahashi}}
\email{takahashi.akira.36m@gmail.com}
\altaffiliation[Present address: ]{Laboratory for Materials and Structures, Institute of Innovative Research, Tokyo Institute of Technology}
\affiliation{Department of Materials Science and Engineering, Kyoto University, Kyoto 606-8501, Japan}
\author{Atsuto \surname{Seko}}
\email{seko@cms.mtl.kyoto-u.ac.jp}
\affiliation{Department of Materials Science and Engineering, Kyoto University, Kyoto 606-8501, Japan}
\affiliation{Center for Elements Strategy Initiative for Structure Materials (ESISM), Kyoto University, Kyoto 606-8501, Japan}
\affiliation{JST, PRESTO, Kawaguchi 332-0012, Japan}
\affiliation{Center for Materials Research by Information Integration, National Institute for Materials Science, Tsukuba 305-0047, Japan}
\author{Isao \surname{Tanaka}}
\affiliation{Department of Materials Science and Engineering, Kyoto University, Kyoto 606-8501, Japan}
\affiliation{Center for Elements Strategy Initiative for Structure Materials (ESISM), Kyoto University, Kyoto 606-8501, Japan}
\affiliation{Center for Materials Research by Information Integration, National Institute for Materials Science, Tsukuba 305-0047, Japan}
\affiliation{Nanostructures Research Laboratory, Japan Fine Ceramics Center, Nagoya 456-8587, Japan}

\date{\today}

\begin{abstract}
Machine-learning interatomic potential (MLIP) has been of growing interest as a useful method to describe the energetics of systems of interest.
In the present study, we examine the accuracy of linearized pairwise MLIPs and angular-dependent MLIPs for 31 elemental metals.
Using all of the optimal MLIPs for 31 elemental metals, we show the robustness of the linearized frameworks, the general trend of the predictive power of MLIPs and the limitation of pairwise MLIPs.
As a result, we obtain accurate MLIPs for all 31 elements using the same linearized framework.
This indicates that the use of numerous descriptors is the most important practical feature for constructing MLIPs with high accuracy.
An accurate MLIP can be constructed using only pairwise descriptors for most non-transition metals, whereas it is very important to consider angular-dependent descriptors when expressing interatomic interactions of transition metals.
\end{abstract}

\maketitle

\section{Introduction}

\begin{figure*}[tbp]
\includegraphics[clip,width=\linewidth]{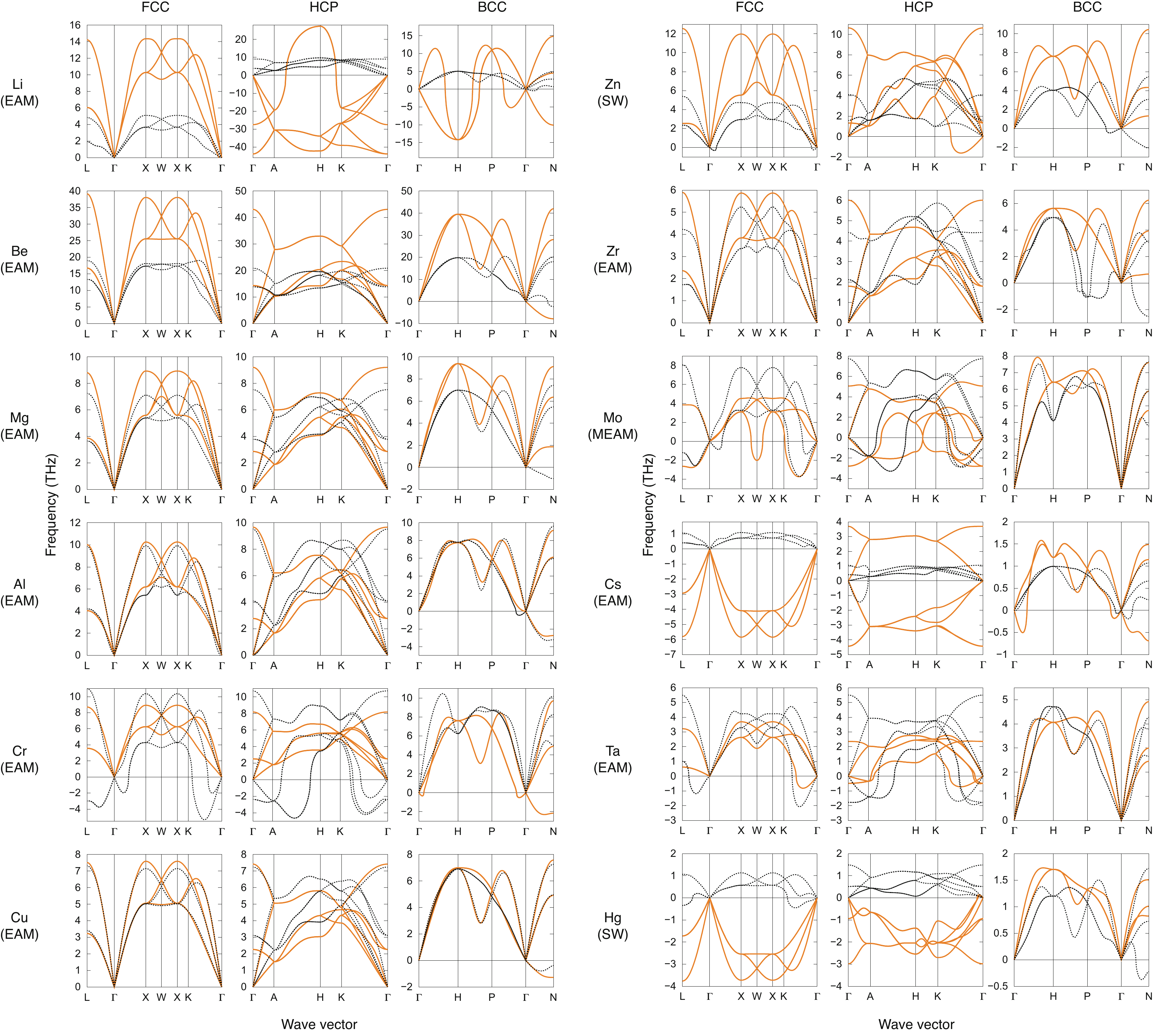}
\caption{
Phonon dispersion curves of fcc, hcp and bcc structures computed using 
EAM (Li)\cite{doi:10.1063/1.1555275}, 
EAM (Be)\cite{0965-0393-21-8-085001}, 
EAM (Mg)\cite{PhysRevB.73.024116}, 
EAM (Al)\cite{PhysRevB.68.024102}, 
EAM (Cr)\cite{0965-0393-21-8-085004}, 
EAM (Cu)\cite{PhysRevB.63.224106}, 
Stillinger--Weber (SW) (Zn)\cite{PhysRevB.88.085309}, 
EAM (Zr)\cite{PhysRevB.69.144113}, 
MEAM (Mo)\cite{PhysRevB.85.214121}, 
EAM (Cs)\cite{doi:10.1063/1.1555275}, 
EAM (Ta)\cite{PhysRevB.69.144113} and 
SW (Hg) potentials\cite{PhysRevB.88.085309}, 
shown by the orange solid lines.
All interatomic potentials are taken from NIST interatomic potential repository\cite{NIST-IPR}.
Phonon dispersion curves computed by DFT calculations are also shown by the black broken lines.
}
\label{Fig-ConvPhonon}
\end{figure*}

A wide variety of conventional interatomic potentials (IPs) have been developed on the basis of prior knowledge of chemical bonds in some systems of interest.
Examples include Lennard--Jones, embedded atom method (EAM), modified EAM (MEAM) and Tersoff potentials.
However, the accuracy and transferability of conventional IPs are often lacking owing to the simplicity of their potential forms.
Figure \ref{Fig-ConvPhonon} shows the phonon dispersion relationships of face-centered cubic (fcc), hexagonal close-packed (hcp) and body-centered cubic (bcc) structures for 12 elemental metals computed using conventional IPs taken from a repository \cite{doi:10.1063/1.1555275,0965-0393-21-8-085001,PhysRevB.73.024116,PhysRevB.68.024102,0965-0393-21-8-085004,PhysRevB.63.224106,PhysRevB.88.085309,PhysRevB.69.144113,PhysRevB.85.214121,NIST-IPR}, along with those computed by density functional theory (DFT) calculation. 
As can be seen in Fig. \ref{Fig-ConvPhonon}, the phonon dispersions of conventional IPs markedly deviate from those obtained by DFT calculations, except for fcc-Al, bcc-Al, fcc-Cu and bcc-Cu.
The lack of accuracy and transferability is associated with the use of an insufficient number of functions for describing interatomic interactions.
Because conventional IPs have an acceptable accuracy in limited simple systems, a robust framework to obtain accurate IPs for a wide range of systems has been desired.

Machine-learning IP (MLIP) based on a large dataset obtained by DFT calculations is beneficial for significantly improving the accuracy and transferability
\cite{
behler2007generalized,
bartok2010gaussian,
behler2011atom,
PhysRevB.90.024101,
PhysRevB.90.104108,
PhysRevB.92.054113,
Thompson2015316,
PhysRevMaterials.1.043603,
PhysRevLett.114.096405,
PhysRevB.95.214302,
doi-10.1063-1.4930541,
PhysRevB.92.045131,
QUA:QUA24836,
doi-10.1137-15M1054183}.
MLIPs represent the total energy of a structure as the sum of the constituent atomic energies, as well as conventional IPs.
The atomic energy of atom $i$ is formulated as
\begin{equation}
E^{(i)} = F\left(d_1^{(i)}, d_2^{(i)}, \cdots, d_{n_{\rm max}}^{(i)} \right),
\label{Eqn-MLIP}
\end{equation}
where $d_n^{(i)}$ denotes a function depending on the neighboring environment of atom $i$.
A major difference between conventional IPs and MLIPs is in the set of functions $\{d_n^{(i)}\}$ and its theoretical background.
In conventional IPs, the set of functions is composed of a few functions expressing atomic interactions and/or physical properties related to the local neighboring environment.
On the other hand, the set of functions in MLIPs is composed of numerous functions, which are generally called ``descriptors''.
The use of numerous functions supports the high accuracy of MLIPs.
In addition, because MLIPs are generally regarded as extensions of conventional IPs, the classification of conventional IPs into the pair potential, pair functional potential, cluster potential and cluster functional potential \cite{carlsson1990beyond} can also be applied to MLIPs in accordance with the type of descriptors.

In our previous studies \cite{PhysRevB.90.024101,PhysRevB.92.054113}, a linearized MLIP, in which pairwise descriptors and their powers are considered, was applied to 12 elemental metals.
In Ref. \onlinecite{takahashi2017conceptual}, we introduced an interpretation of the MLIP based on the framework of EAM and MEAM potentials, which gives both conceptual and practical reasons for the high accuracy of MLIPs.
On the basis of the interpretation, we also proposed more general linearized frameworks and obtained an MLIP with high accuracy for elemental Ti.

\begin{figure}[tbp]
\includegraphics[clip,width=0.9\linewidth]{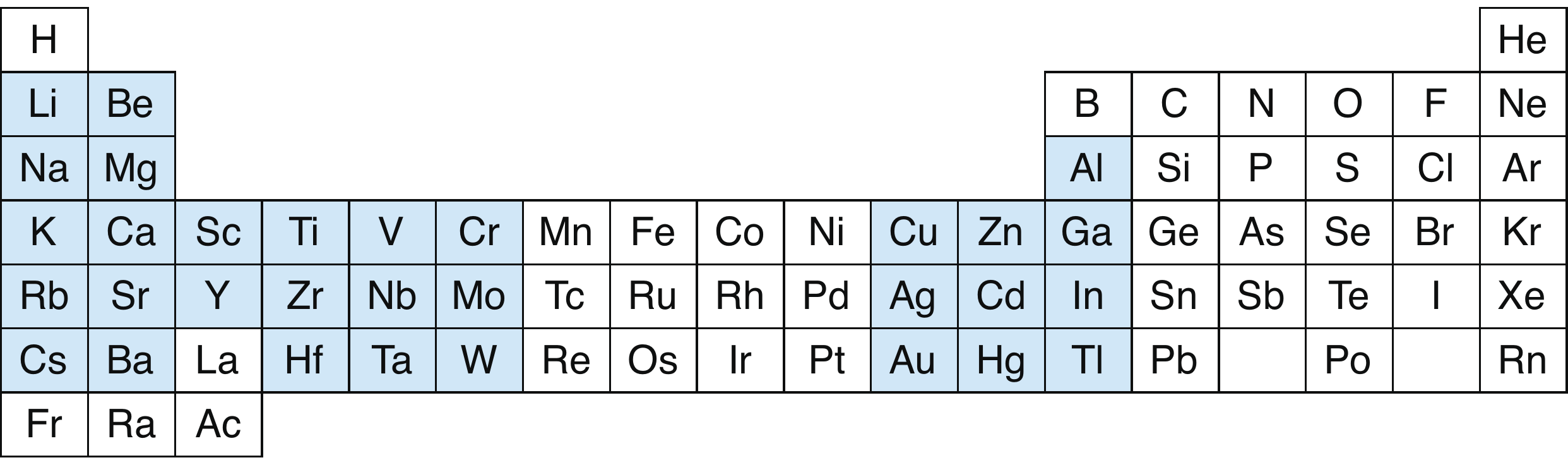}
\caption{
Target 31 elemental metals used to construct MLIPs in this study, shown by the blue shaded blocks.
}
\label{Fig-PeriodicTable}
\end{figure}

In this study, we adopt all of the linearized frameworks to construct MLIPs for the 31 elemental metals shown in Fig. \ref{Fig-PeriodicTable}.
Using all of the optimal MLIPs for the 31 elemental metals, we show the robustness of the linearized frameworks.
Also, the general trend of the predictive power of MLIPs and the limitation of pairwise MLIPs are demonstrated, which can only be investigated by applying MLIP frameworks to a variety of systems.
There have been no studies in which an MLIP framework has been systematically applied to a wide range of elemental metals or other systems.
Section \ref{Sec-Result} reports the predictive power of MLIPs for elemental Cu, Al and Cr.
Predictions for the other elemental metals are given in the Appendix.

%

\section{Methodology}
\subsection{Modeling of atomic energy}
The modeling of atomic energy begins with Eqn. (\ref{Eqn-MLIP}).
We employ two types of descriptors: pairwise and angular-dependent descriptors.
Pairwise descriptors are given by the sum of pairwise functions taken over neighboring atoms.
They are written as
\begin{equation}
\label{nanoinfo:EqnGRDF}
b_{n0}^{(i)} = \sum_j f_n (r_{ij}),
\end{equation}
where $f_n (r_{ij})$ denotes a pairwise function of distance $r_{ij}$ between atoms $i$ and $j$.
Here smooth Gaussian-type radial functions are adopted as $f_n(r)$, expressed as
\begin{equation}
f_n(r) = \exp \left[-p_n (r-q_n)^2\right]f_c(r),
\end{equation}
where $p_n$ and $q_n$ denote given parameters.
$f_c(r)$ is a smooth cutoff function that is equal to zero at a distance exceeding a cutoff radius $r_c$.
We use a cosine-based cutoff function expressed as
\begin{eqnarray}
f_c(r) = \left\{
\begin{aligned}
& \frac{1}{2} \left[ \cos \left( \pi \frac{r}{r_c} \right) + 1\right] & (r \le r_c)\\
& 0 & (r > r_c)
\end{aligned}
\right . .
\end{eqnarray}

We also use angular-dependent descriptors that can include both the radial and angular dependences of the neighboring atomic distribution.
One such descriptor is the angular Fourier series (AFS)\cite{bartok2013representing} given as 
\begin{equation}
b_{nl}^{(i)} = \sum_{j,k} f_n(r_{ij})f_n(r_{ik}) \cos (l \theta_{ijk}),
\end{equation}
where $\theta_{ijk}$ denotes the bond angle between atom $i$ and its neighboring two atoms.
As well as pairwise descriptors, Gaussian-type radial functions are employed as $f_n$ in the AFS.
The AFS corresponds to a set of rotationally invariant descriptors derived from products of radial and spherical harmonic functions\cite{bartok2013representing} that can be found in the literature\cite{carlsson1990beyond}.

Although artificial neural network and Gaussian process black-box models have been used as a function $F$ in most of the literature, we introduce simple polynomial functions as a function $F$.
The features of such linearized MLIP frameworks are as follows.
(1) The relationship with conventional IPs can be simply considered because a black-box model is not employed.
(2) The forces acting on atoms and stress tensors are easily included in the training data in a straightforward manner because they are also expressed by linear equations\cite{PhysRevB.92.054113}.
(3) A sparse representation of the MLIP is obtained using the Lasso or related techniques, which decreases the computational cost for the energy and forces\cite{PhysRevB.90.024101}. 
(4) Regression coefficients are simply estimated using a fast standard machine-learning technique.
(5) The number of regression coefficients and the computational cost for the energy and forces are formally independent of the number of training data.

Herein, the atomic energy is modeled using three combinations of descriptor types and polynomial types.
In the first approximation, only pairwise descriptors $b_{n0}^{(i)}$, their squares and their cubes are considered (hereafter called pairwise-MLIP1).
This is identical to the model of the atomic energy introduced in Refs. \onlinecite{PhysRevB.90.024101,PhysRevB.92.054113}.
The atomic energy is expressed as
\begin{eqnarray}
E^{(i)} & = & w_0 + \displaystyle\sum_n w_{n0} b_{n0}^{(i)} + \sum_n w_{n0,n0} b_{n0}^{(i)}b_{n0}^{(i)} \nonumber \\
& & + \displaystyle \sum_n w_{n0,n0,n0} b_{n0}^{(i)}b_{n0}^{(i)}b_{n0}^{(i)}, 
\end{eqnarray}
where $w_0$, $w_{n0}$, $w_{n0,n0}$ and $w_{n0,n0,n0}$ denote the regression coefficients. 
This model is regarded as a pair functional potential, which is in the same class of potentials as EAM potentials.
In this model, many-body atomic interactions are effectively modeled by powers of the pairwise descriptors.

The second approximation is a third-order polynomial approximation with only pairwise descriptors, which adopts a more general polynomial function than that used in the first approximation (hereafter called pairwise-MLIP2).
The second approximation describes the atomic energy as
\begin{eqnarray}
E^{(i)} & = & w_0 + \displaystyle\sum_n w_{n0} b_{n0}^{(i)} + \sum_{n,n'} w_{n0,n'0} b_{n0}^{(i)}b_{n'0}^{(i)} \nonumber \\
& & + \displaystyle \sum_{n,n',n''} w_{n0,n'0,n''0} b_{n0}^{(i)}b_{n'0}^{(i)}b_{n''0}^{(i)},
\end{eqnarray}
where cross terms of descriptors are included in addition to the powers of descriptors.
This model is also classified into a pair functional potential.

The third approximation is a second-order polynomial approximation with pairwise and angular-dependent descriptors, which is classified as a cluster functional potential (hereafter called an angular-dependent MLIP).
It should be important to include angular-dependent terms to describe the atomic energy accurately (e.g., \cite{takahashi2017conceptual}).
The atomic energy is written as
\begin{equation}
E^{(i)} = w_0 + \displaystyle\sum_{nl} w_{nl} b_{nl}^{(i)} + \sum_{n,l,n',l'} w_{nl,n'l'} b_{nl}^{(i)}b_{n'l'}^{(i)} .
\end{equation}
Here we consider all descriptors up to the maximum value of $l_{\rm max} = 10$.

\subsection{Estimation of regression coefficients}
The vector $\bm{w}$ composed of all the regression coefficients is estimated by regression, which is a machine learning method to estimate the relationship between the predictor and observation variables using a training dataset.
For the training data, the energy, the forces acting on atoms and the stress tensor computed by DFT calculations can be used as the observations in the regression process since they are all expressed by linear equations with the same regression coefficients \cite{PhysRevB.92.054113}.
Here linear ridge regression \cite{hastieelements} is employed to estimate the regression coefficients. 
Linear ridge regression shrinks the regression coefficients by imposing a penalty.
The ridge coefficients minimize the penalized residual sum of squares and are expressed as
\begin{equation}
L(\bm{w}) = ||\bm{X}\bm{w} - \bm{y}||^2_2 + \lambda ||\bm{w}||^2_2, 
\end{equation}
where $\bm{X}$ and $\bm{y}$ denote the predictor matrix and observation vector, respectively, corresponding to the training data.
Parameter $\lambda$, which is called the regularization parameter, controls the magnitude of the penalty.
This is referred to as L2 regularization.
The regression coefficients can easily be estimated while avoiding the well-known multicollinearity problem that occurs in the ordinary least-squares method.

In all three models of the atomic energy, the complexity of the model or the number of regression coefficients is controlled by only the number of Gaussian radial functions $f_n$.
Therefore, the number of Gaussian functions and the parameters of the Gaussian functions are optimized systematically.
Parameter $p_n$ is given as a single value, whereas parameter $q_n$ is given as each value of an arithmetic sequence defined by a given maximum value of $q_n$ and the number of Gaussian functions.
The cutoff radius $r_c$ is also optimized using the convergence of the prediction error for the energy.
Its definition will be given later.
Tables \ref{Table-Optimal1}, \ref{Table-Optimal2} and \ref{Table-Optimal3} in the Appendix respectively summarize optimal values of the cutoff radius, the number of Gaussian functions and the parameters of the Gaussian functions for pairwise-MLIP1, pairwise-MLIP2 and the angular-dependent MLIP.

\subsection{DFT calculation}
Training and test datasets were generated from DFT calculations.
The test dataset was used to examine the predictive power for structures that are not included in the training dataset.
For each elemental metal, 2700 and 300 configurations were generated for the training and test datasets, respectively.
Therefore, a total of 93,000 DFT calculations were performed.
The datasets include structures generated by isotropic expansions, random expansions, random distortions and random displacements of ideal fcc, bcc, hcp, simple cubic (sc), $\omega$ and $\beta$-tin structures, in which the atomic positions and lattice constants were fully optimized.
These configurations were made using supercells constructed by the $2\times2\times2$, $3\times3\times3$, $3\times3\times3$, $4\times4\times4$, $3\times3\times3$ and $2\times2\times2$ expansions of the conventional unit cells for fcc, bcc, hcp, sc, $\omega$ and $\beta$-tin structures, respectively.
Therefore, they are composed of 32, 54, 54, 64, 81 and 32 atoms, respectively.

For the total of 3000 configurations for each elemental metal, DFT calculations were performed using the plane-wave basis projector augmented wave (PAW) method \cite{PAW1} within the Perdew--Burke--Ernzerhof exchange-correlation functional \cite{GGA:PBE96} as implemented in the \textsc{VASP} code \cite{VASP1,VASP2,PAW2}.
The cutoff energy was set to 400 eV.
The total energies converged to less than 10$^{-3}$ meV/supercell.
The atomic positions and lattice constants were optimized for the ideal structures until the residual forces were less than 10$^{-3}$ eV/\AA.

\begin{figure}[tbp]
\includegraphics[clip,width=0.95\linewidth]{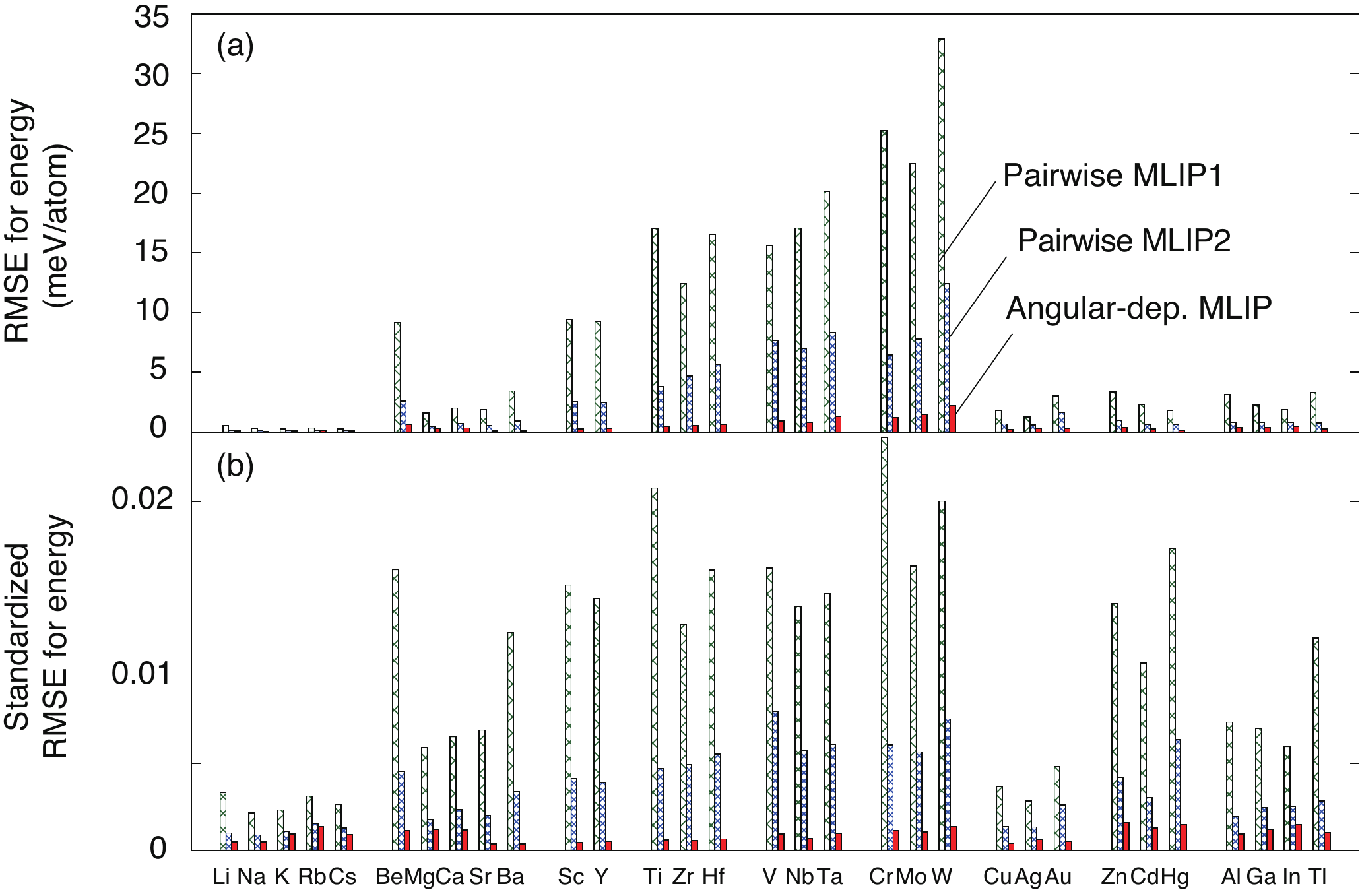}
\caption{
(a) RMSE and (b) standardized RMSE for energy of MLIPs for 31 elements.
}
\label{Fig-RmseAll}
\end{figure}

\begin{table*}[htbp]
\begin{ruledtabular}
\caption{
RMSEs for energy, force and elastic constant averaged over all 31 elemental metals, non-transition metals and transition metals.
The average differences of the phonon DOS between DFT and the MLIPs are also shown.
The units for energy $E$, force $f$ and elastic constant $C$ are meV/atom, eV/\AA\ and GPa, respectively.
}
\label{Table-Average}
\begin{tabular}{c|cccc|cccc|cccc}
& \multicolumn{4}{c|}{All elemental metals} & \multicolumn{4}{c|}{Non-transition metals} &
  \multicolumn{4}{c}{Transition metals} \\
& $E$ & $f$ & $C$ & $\Delta_{\rm ph}$ & $E$ & $f$ & $C$ & $\Delta_{\rm ph}$ & $E$ & $f$ & $C$ & $\Delta_{\rm ph}$ \\
\hline
Pairwise-MLIP1    & 7.8 & 0.057 & 46.8 & 0.25 &  2.1 & 0.014 & 15.1 & 0.18 & 18.0 & 0.135 & 104.3 & 0.36 \\
Pairwise-MLIP2    & 2.7 & 0.037 & 21.3 & 0.17 &  0.7 & 0.008 &  9.7 & 0.13 &  6.3 & 0.090 &  42.4 & 0.24 \\
Ang.-dep. MLIP    & 0.5 & 0.011 & 11.7 & 0.11 &  0.3 & 0.002 &  8.7 & 0.09 &  0.9 & 0.027 &  17.1 & 0.15 \\
\end{tabular}
\end{ruledtabular}
\end{table*}

\section{Results and discussion}
\label{Sec-Result}
\subsection{Trend of predictive power}
\label{Sec-Trend}

For each of the 31 elemental metals, the optimal pairwise-MLIP1, pairwise-MLIP2 and angular-dependent MLIP were constructed following the above procedure.
The predictive power of each MLIP was estimated using the root-mean-square error (RMSE) between the observation property predicted by the DFT calculation and that predicted by the MLIP for the test data. 
Here, the energy, the force acting on atoms and the elastic constant were used as the observation properties.
The predictive power was also estimated by the difference in the phonon density of states (DOS) between the DFT calculation and the MLIP.
The difference in the phonon DOS, $\Delta_{\rm ph}$, is defined using the L1 norm of the difference in the phonon DOS histograms as
\begin{equation}
\Delta_{\rm ph} = \frac{\Delta h}{2} \sum_i \left| D^{\rm DFT}(\omega_i) - D^{\rm MLIP}(\omega_i) \right|,
\end{equation}
where $D(\omega_i)$ and $\Delta h$ denote the DOS for the $i$th bin normalized by the degree of freedom and the bin width, respectively.

Figure \ref{Fig-RmseAll} (a) and Tables \ref{Table-Optimal1}--\ref{Table-Optimal3} show the RMSE for the energy of the optimal MLIPs for the 31 elemental metals.
Tables \ref{Table-Optimal1}--\ref{Table-Optimal3} also show the RMSE for the force of the optimal MLIPs for the 31 elemental metals.
Table \ref{Table-Average} shows the RMSEs for the energy and force of the optimal MLIPs averaged over the 31 elemental metals, the non-transition metals and the transition metals.
The average RMSEs for the energy and force for pairwise-MLIP1 are 7.8 meV/atom and 0.057 eV/\AA, respectively.
Although these are large values, pairwise-MLIP1 shows a small RMSE for non-transition metal elements except for Be, which leads to the result that the RMSEs averaged over the non-transition metal elements are 2.1 meV/atom and 0.014 eV/\AA.
On the other hand, pairwise-MLIP1 shows large average RMSEs of 18.0 meV/atom and 0.135 eV/\AA\ for the transition metals. 

The average RMSEs decrease when considering the cross terms of pairwise descriptors for all the elemental metals.
As a result, the average RMSEs of pairwise-MLIP2 are 2.7 meV/atom and 0.037 eV/\AA, which means that many-body atomic interactions are more effectively included in pairwise-MLIP2 than in pairwise-MLIP1.
For most of the non-transition metals, the average RMSEs of pairwise-MLIP2 are 0.7 meV/atom and 0.008 eV/\AA. 
In other words, an accurate MLIP can be constructed using only pairwise descriptors for most of the non-transition metals.
This indicates that the lack of accuracy of conventional pairwise (pair functional) IPs such as EAM potentials can be attributed to the use of an insufficient number of descriptors in the conventional IPs.
In contrast, pairwise-MLIP2 for the transition metals shows large average RMSEs of 6.3 meV/atom and 0.090 eV/\AA.

By considering pairwise and angular-dependent descriptors and their cross terms, the average RMSEs for the energy and force are reduced to 0.5 meV/atom and 0.011 eV/\AA, respectively.
Regarding the transition metals, the average RMSEs for the energy and force significantly decrease from 6.3 to 0.9 meV/atom and from 0.090 to 0.027 eV/\AA, respectively.
This means that angular-dependent descriptors are essential for expressing the interatomic interactions of elemental metals very accurately, particularly in transition metals.

\begin{figure}[tbp]
\includegraphics[clip,width=0.9\linewidth]{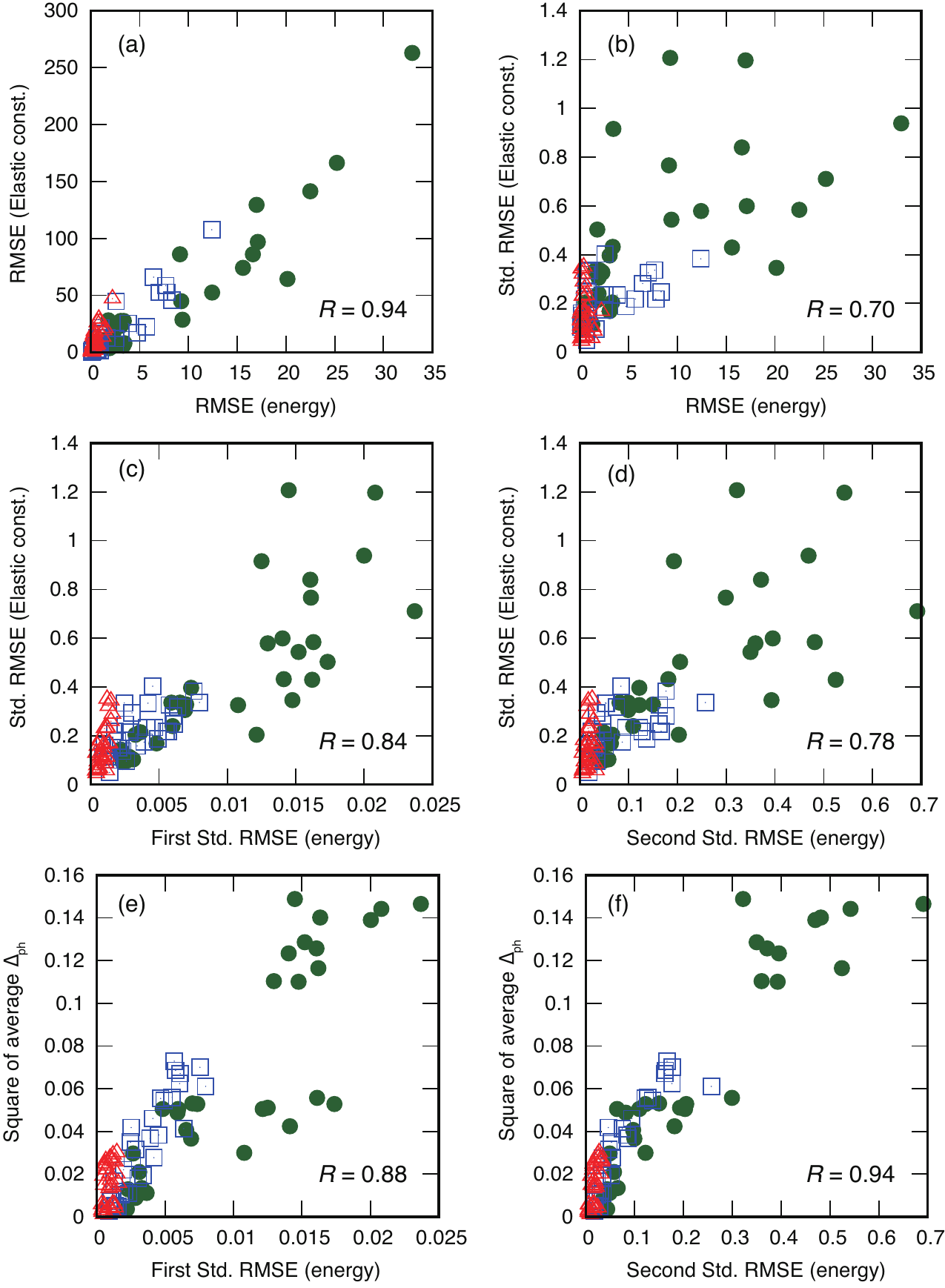}
\caption{
Correlations (a) between RMSE for energy and RMSE for elastic constant,
(b) between RMSE for energy and standardized RMSE for elastic constant,
(c) between first standardized RMSE for energy and standardized RMSE for elastic constant,
(d) between second standardized RMSE for energy and standardized RMSE for elastic constant,
(e) between first standardized RMSE for energy and square of average difference in phonon DOS
and (f) between second standardized RMSE for energy and square of average difference in phonon DOS.
}
\label{Fig-CorrRMSE}
\end{figure}

Then, the RMSE for the elastic constant and the difference in the phonon DOS are demonstrated.
Table \ref{Table-Average} shows the RMSEs for the elastic constant and the difference in the phonon DOS averaged over the 31 elemental metals, the non-transition metals and the transition metals. 
Both the RMSE for the elastic constant and the difference in the phonon DOS have the same tendency as the average RMSEs for the energy and force.

Figure \ref{Fig-CorrRMSE} (a) shows the relationship between the RMSEs for the energy and the elastic constant.
They have a strong correlation with a correlation coefficient of $R=0.94$.
Therefore, the RMSE for the energy is regarded as a rough estimator of the prediction error of the elastic constants.
However, the required accuracy for the elastic constants depends on their absolute values.
In other words, the RMSE depends on the intrinsic energy scale of the element, corresponding to the energy distribution for structures.
Therefore, it is sometimes preferable to use the standardized RMSE for the elastic constant when estimating the predictive power of the elastic constant and when comparing the performance of MLIPs in different systems.
Here, we employ the standardized RMSE for the elastic constant divided by the bulk modulus.
Figure \ref{Fig-CorrRMSE} (b) shows the relationship between the RMSE for the energy and the standardized RMSE for the elastic constant.
This indicates that the correlation between the RMSEs becomes weak upon standardizing the RMSE for the elastic constant.

In a natural sense, the RMSE for the energy should also be standardized to compare it with the standardized RMSE for the elastic constant.
We standardize the RMSE for the energy in two ways.
One is to use the standard deviation computed from the energies of all structures.
The other is to use the standard deviation computed from the energies of a set of structures excluding outlier structures generated by isotropic expansion. 
To capture the local energy distribution around ideal structures, we split all structures into six sets of structures generated from the ideal fcc, hcp, bcc, sc, $\omega$ and $\beta$-tin structures.
Then, the average of the standard deviations for the six structure sets was used to standardize the RMSE for the energy.
Figures \ref{Fig-CorrRMSE} (c) and (d) show the relationships between the first and second standardized RMSEs for the energy and the standardized RMSE for the elastic constant, respectively.
As can be seen in Fig. \ref{Fig-CorrRMSE}, the first standardized method is better because a stronger correlation with the standardized RMSE for the elastic constant is found.
Regarding the difference in the phonon DOS, both the standardized RMSEs for the energy have a strong correlation with the difference in the phonon DOS as shown in Figs. \ref{Fig-CorrRMSE} (e) and (f).
Therefore, the standardized RMSEs for energy can be an estimator of the prediction error of phonon properties.
Consequently, the RMSE for the energy is logically acceptable as a quantity to be minimized to optimize MLIPs.

Figure \ref{Fig-RmseAll} (b) shows the first standardized RMSE for the energy of the optimal MLIPs for the 31 elemental metals.
All transition metals exhibit similar behavior, and the standardized RMSEs of many non-transition metals are close to those of the transition metals, which is a different trend from that found in the RMSE.
In Be, Ba, Tl and the elements of group 12, the standardized RMSE for the energy is very large in spite of the small RMSE for the energy.
This means that the inconsistency of the physical properties between the MLIP and the DFT calculation may be large despite the RMSE for the energy being small.

\subsection{Copper}
\label{Sec-Cu}

\begin{figure}[tbp]
\includegraphics[clip,width=\linewidth]{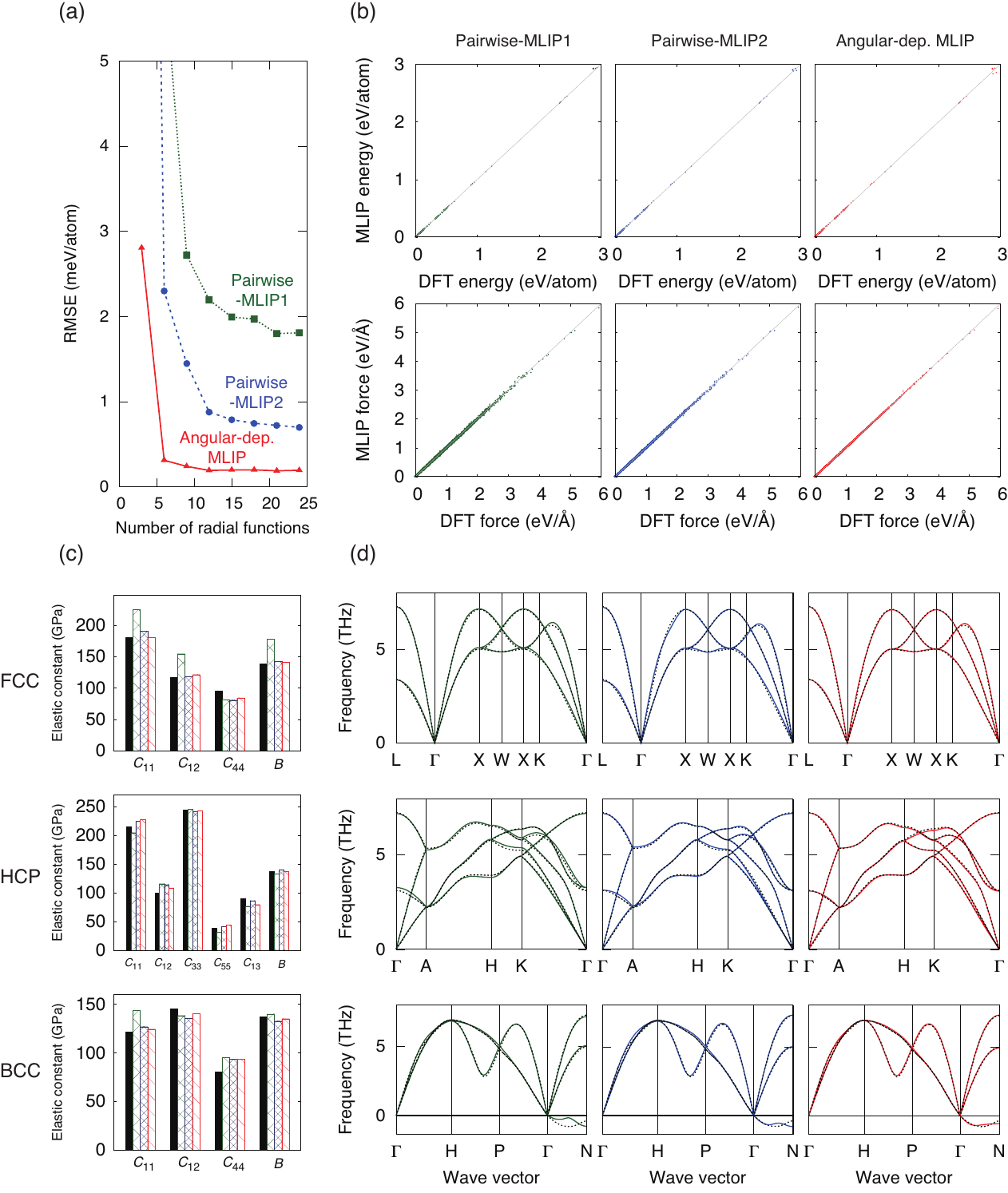}
\caption{
(a) Dependence of RMSE for energy on the number of Gaussian radial functions for Cu.
The converged RMSEs correspond to the RMSEs of the optimal MLIPs.
(b) Distributions of the energy and absolute force acting on atoms predicted by DFT and the MLIPs.
(c) Elastic constants of fcc, hcp and bcc Cu predicted by the optimal MLIPs.
(d) Phonon dispersion curves of fcc, hcp and bcc Cu predicted by the optimal MLIPs.
}
\label{Fig-Cu}
\end{figure}

Here we demonstrate applications of linearized MLIPs to elemental Cu, for which the EAM potential has good predictive power as shown in Fig. \ref{Fig-ConvPhonon}.
Figure \ref{Fig-Cu} (a) shows the dependence of the RMSE on the number of Gaussian radial functions for Cu.
The RMSEs of the optimal pairwise-MLIP1, pairwise-MLIP2 and angular-dependent MLIP are 1.88, 0.70 and 0.19 meV/atom, respectively.
Even though angular-dependent descriptors are excluded, an MLIP with an RMSE of less than 1 meV/atom can be obtained.
Figure \ref{Fig-Cu} (b) shows the distributions of the energy and absolute force acting on atoms predicted by DFT and the MLIPs. 
There is little difference between the DFT and MLIP energies and forces.
Figures \ref{Fig-Cu} (c) and (d) show the elastic constants and phonon dispersion curves of fcc, hcp and bcc Cu predicted by the optimal pairwise-MLIP1, pairwise-MLIP2 and angular-dependent MLIP, respectively, along with those predicted by the DFT calculation.
The phonon dispersion curves predicted by the three MLIPs are almost identical to those predicted by the DFT calculation.
Although conventional EAM potentials are known to predict properties with acceptable accuracy as shown in Fig. \ref{Fig-ConvPhonon}, the phonon dispersion curves of the MLIPs are more consistent with those of the DFT calculations than those of the EAM potential.
The elastic constants of pairwise-MLIP1 slightly differ from those of the DFT calculation, but the elastic constants are predicted accurately by all the MLIPs as well as the phonon dispersion curves.
The same tendency of the predictive power is recognized for other elements of groups 1, 2, 11 except for Be, although good conventional IPs have not been developed for such elemental metals, for example, Li, Mg and Cs, as shown in Fig. \ref{Fig-ConvPhonon}.

\subsection{Aluminum}
\label{Sec-Al}

\begin{figure}[tbp]
\includegraphics[clip,width=\linewidth]{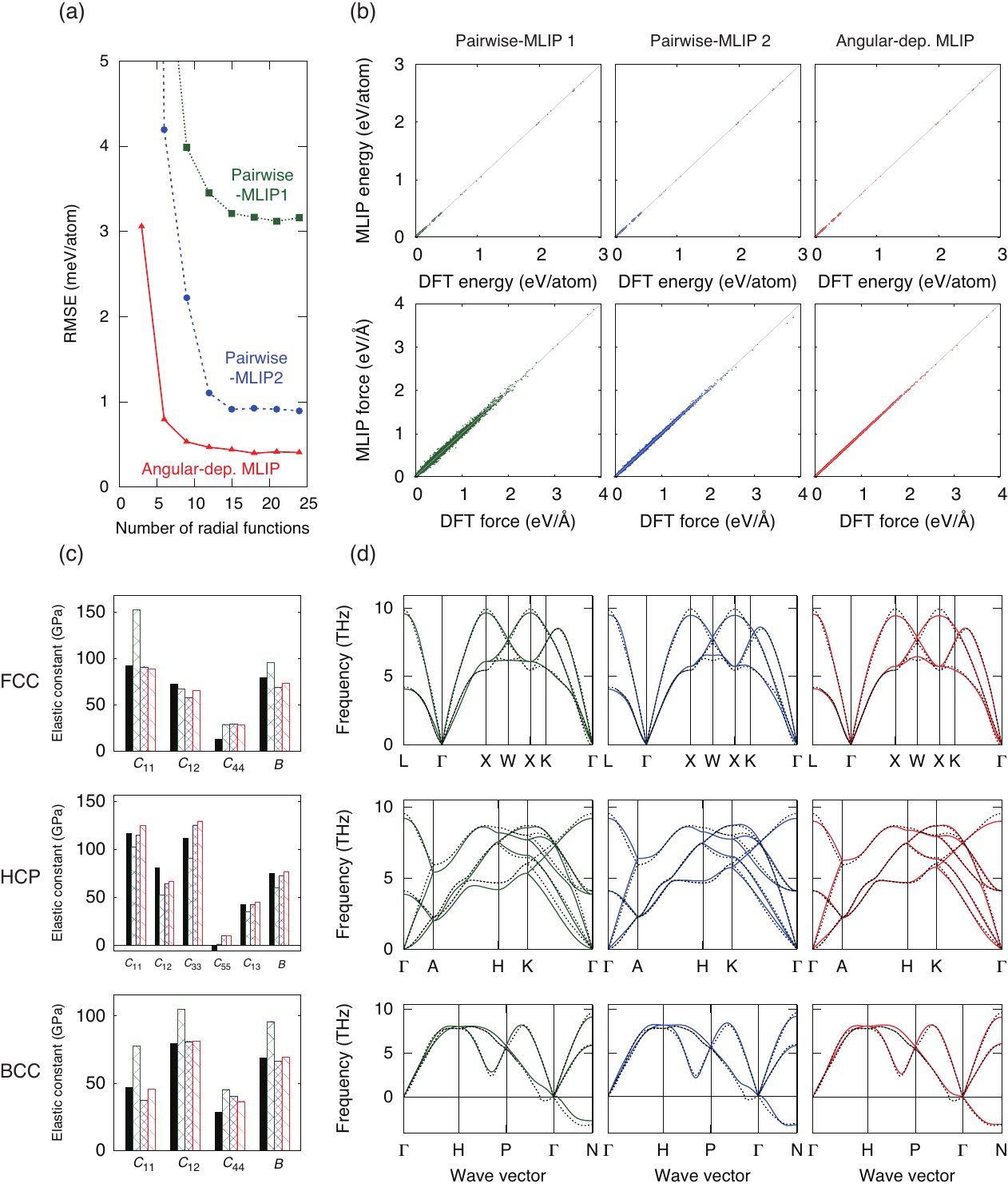}
\caption{
(a) Dependence of RMSE for energy on the number of Gaussian radial functions for Al.
The converged RMSEs correspond to the RMSEs of the optimal MLIPs.
(b) Distributions of the energy and absolute force acting on atoms predicted by DFT and the MLIPs.
(c) Elastic constants of fcc, hcp and bcc Al predicted by the optimal MLIPs.
(d) Phonon dispersion curves of fcc, hcp and bcc Al predicted by the optimal MLIPs.
}
\label{Fig-Al}
\end{figure}

In elemental Al, the EAM potential well predicts the phonon properties of fcc and bcc structures as shown in Fig. \ref{Fig-ConvPhonon}.
However, its predictive power for Al is slightly poorer than that for Cu.
Figure \ref{Fig-Al} (a) shows the dependence of the RMSE on the number of Gaussian radial functions for Al.
The RMSEs of the optimal pairwise-MLIP1, pairwise-MLIP2 and angular-dependent MLIP are 3.67, 0.89 and 0.40 meV/atom, respectively.
Although the RMSE of pairwise-MLIP1 is approximately twice that of Cu, an MLIP with an RMSE of less than 1 meV/atom is obtained without considering angular-dependent descriptors.
Figure \ref{Fig-Al} (b) shows the distributions of the energy and absolute force acting on atoms predicted by DFT and the MLIPs.
As well as for Cu, the DFT and MLIP energies are very close for all structures.
The MLIP forces slightly deviate from the DFT forces in pairwise-MLIP1, whereas the MLIP forces are almost identical to the DFT forces in pairwise-MLIP2 and the angular-dependent MLIP.
Figures \ref{Fig-Al} (c) and (d) show the elastic constants and phonon dispersion curves of fcc, hcp and bcc Al predicted by the optimal pairwise-MLIP1, pairwise-MLIP2 and angular-dependent MLIP, respectively, along with those predicted by the DFT calculation.
Although the phonon dispersion curves of pairwise-MLIP1 slightly deviate from those of the DFT calculation, the consistency with the DFT calculation is better than that of the EAM potential.
The phonon dispersion curves of pairwise-MLIP2 and the angular-dependent MLIP are almost consistent with those of the DFT calculation.
As well as the phonon dispersion, the elastic constants predicted by pairwise-MLIP1 are inconsistent with those predicted by the DFT calculation.
The inconsistency of the elastic constants is avoided by including cross terms of pairwise descriptors.
Pairwise-MLIP2 and the angular-dependent MLIP predict the elastic constants within an acceptable accuracy.
A similar trend can also be observed for the elements of groups 12 and 13 except for Hg.

\subsection{Chromium}
\label{Sec-Cr}

\begin{figure}[tbp]
\includegraphics[clip,width=\linewidth]{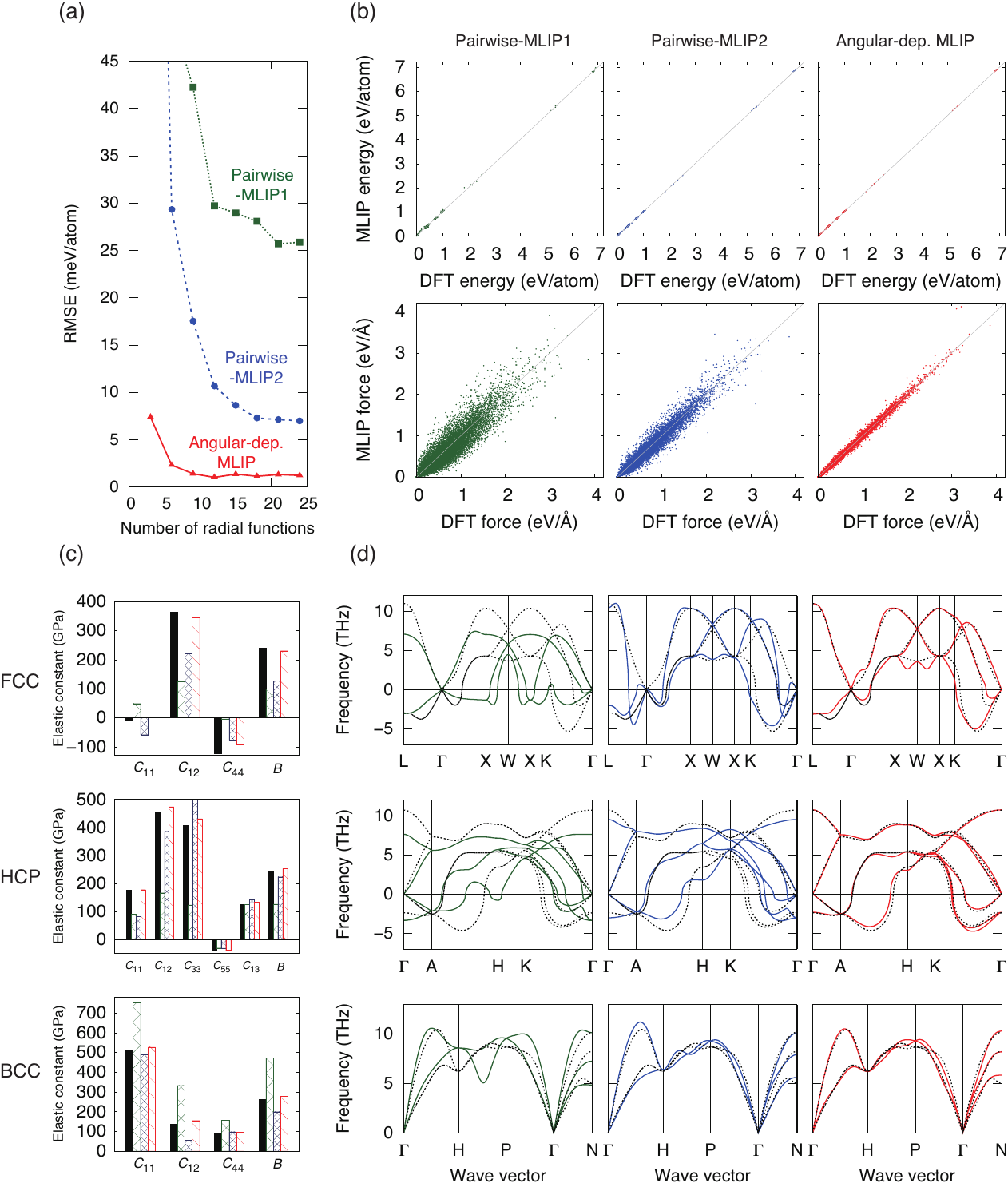}
\caption{
(a) Dependence of RMSE for energy on the number of Gaussian radial functions for Cr.
The converged RMSEs correspond to the RMSEs of the optimal MLIPs.
(b) Distributions of the energy and absolute force acting on atoms predicted by DFT and the MLIPs.
(c) Elastic constants of fcc, hcp and bcc Cr predicted by the optimal MLIPs.
(d) Phonon dispersion curves of fcc, hcp and bcc Cr predicted by the optimal MLIPs.
}
\label{Fig-Cr}
\end{figure}

The third demonstration of applications is for elemental Cr, for which the EAM potential predicts phonon properties very poorly as shown in Fig. \ref{Fig-ConvPhonon}.
Figure \ref{Fig-Cr} (a) shows the dependence of the RMSE on the number of Gaussian radial functions for the elemental Cr.
The RMSEs of the optimal pairwise-MLIP1, pairwise-MLIP2 and angular-dependent MLIP are 25.69, 6.99 and 1.21 meV/atom, respectively.
When angular-dependent descriptors are not considered, the MLIPs show a large RMSE.
On the other hand, angular-dependent descriptors improve the accuracy of the MLIPs.
Figure \ref{Fig-Cr} (b) shows the distributions of the energy and absolute force acting on atoms predicted by DFT and the MLIPs.
The DFT and MLIP energies for all structures are very close as well as for Cu and Al.
However, the MLIP forces markedly differ from the DFT forces in the pairwise MLIPs.
Angular-dependent descriptors improve the prediction of the forces.
Figures \ref{Fig-Cr} (c) and (d) respectively show the elastic constants and phonon dispersion curves of fcc, hcp and bcc Cr predicted by the optimal pairwise-MLIP1, pairwise-MLIP2 and angular-dependent MLIP.
The phonon dispersion curves predicted by the pairwise MLIPs markedly deviate from those predicted by the DFT calculation, whereas the phonon dispersion curves predicted by the angular-dependent MLIP are almost identical to the DFT phonon dispersion curves.
The elastic constants show a similar tendency to the phonon dispersion curves.
The elastic constants of pairwise-MLIP1 are completely different from those of the DFT calculation.
The cross terms of descriptors and angular-dependent descriptors improve the prediction of the elastic constants.
For the other transition-metal elements of groups 3--6, Be and Hg, the same tendency of the predictive power is recognized.
In such elemental metals, angular-dependent descriptors are essential for accurately describing the atomic energy.

\section{Conclusion}
We construct pairwise and angular-dependent linearized MLIPs for 31 elemental metals.
We obtain very accurate MLIPs for all 31 elements using a linearized framework with a simple polynomial function, which shows the robustness of the linearized MLIP framework for elemental metals.
In addition, the fact that MLIPs with high accuracy are obtained using simple polynomial functions for all 31 elements shows that the use of numerous descriptors is the most important practical feature for constructing MLIPs with high accuracy.

Moreover, a general trend of the predictive power of MLIPs and the limitation of pairwise MLIPs are demonstrated.
In non-transition metals except for Be and Hg, accurate MLIPs are obtained using only pairwise descriptors.
The prediction error for the energy averaged over the non-transition metals is 0.7 meV/atom.
This indicates that the lack of accuracy of conventional pairwise (pair functional) IPs such as EAM potentials can be attributed to the use of an insufficient number of descriptors in the conventional IPs.
Meanwhile, angular-dependent descriptors further increases the accuracy of MLIPs, particularly for transition metals.
By considering angular-dependent descriptors, the prediction error for the energy averaged over the transition metals decreases from 6.3 to 0.9 meV/atom.
Systematic angular-dependent descriptors are essential to construct accurate MLIPs for the transition metals and outlier non-transition metals such as Be and Hg.

\begin{acknowledgments}
This study was supported by PRESTO from JST, a Grant-in-Aid for Scientific Research (B) (Grant No. 15H04116) from JSPS and a Grant-in-Aid for Scientific Research on Innovative Areas ``Nano Informatics'' (Grant No. 25106005) from JSPS. 
AT was supported by a Grant-in-Aid for JSPS Research Fellows (Grant No. 15J07315) from JSPS.
\end{acknowledgments}

\appendix
\section{Optimal parameters for MLIPs}
Table \ref{Table-Optimal1} shows the optimal cutoff radius $R_c$, the optimal number of Gaussian radial functions $N_{\rm rad}$ and the parameters for identifying the set of Gaussian radial functions $p$ and $q_{\rm max}$ to construct pairwise-MLIP1.
The RMSEs for the energy and force are also shown in Table \ref{Table-Optimal1}.
The optimal parameters, the number of Gaussian radial functions and the RMSEs for pairwise-MLIP2 and the angular-dependent MLIP are summarized in Tables \ref{Table-Optimal2} and \ref{Table-Optimal3}, respectively.

\begingroup
\squeezetable
\begin{table}[tbp]
\begin{ruledtabular}
\caption{
Optimal cutoff radii, optimal number of Gaussian radial functions, and optimal parameters for identifying the set of Gaussian radial functions and the RMSEs of optimized pairwise-MLIP1 for the 31 elemental metals.
}
\label{Table-Optimal1}
\begin{tabular}{ccccccc}
Element & $R_c$ & $N_{\rm rad}$& $p$ & $q_{\rm max}$ & RMSE (energy) & RMSE (force) \\
& (\AA) & & (\AA$^{-2}$) & (\AA) & (meV/atom) & (eV/\AA) \\
\hline
Li & 13 & 24 & 0.5 & 10.0 & 0.54 & 0.004 \\
Na & 14 & 18 & 0.5 & 12.0 & 0.35 & 0.001 \\
K & 14 & 18 & 0.5 & 11.0 & 0.67 & 0.002 \\
Rb & 12 & 21 & 0.5 & 12.0 & 0.81 & 0.002 \\
Cs & 14 & 15 & 0.5 & 12.0 & 0.80 & 0.002 \\
\hline
Be & 14 & 24 & 2.0 & 12.0 & 9.39 & 0.051 \\
Mg & 7 & 24 & 1.0 & 11.0 & 1.60 & 0.011 \\
Ca & 8 & 24 & 0.5 & 12.0 & 2.05 & 0.016 \\
Sr & 9 & 15 & 0.5 & 11.0 & 1.91 & 0.014 \\
Ba & 9 & 18 & 0.5 & 12.0 & 3.79 & 0.018 \\
\hline
Sc & 7 & 21 & 1.0 & 11.0 & 9.80 & 0.068 \\
Y & 8 & 21 & 1.0 & 10.0 & 9.25 & 0.060 \\
\hline
Ti & 6 & 21 & 1.5 & 10.0 & 17.29 & 0.133 \\
Zr & 7 & 18 & 0.5 & 11.0 & 12.63 & 0.112 \\
Hf & 7 & 18 & 1.0 & 9.0 & 16.56 & 0.139 \\
\hline
V & 12 & 24 & 1.5 & 7.0 & 16.16 & 0.132 \\
Nb & 8 & 24 & 1.5 & 9.0 & 17.08 & 0.139 \\
Ta & 14 & 21 & 2.0 & 8.0 & 20.16 & 0.161 \\
\hline
Cr & 6 & 21 & 1.5 & 7.0 & 25.69 & 0.161 \\
Mo & 14 & 24 & 1.0 & 7.0 & 22.77 & 0.166 \\
W & 9 & 24 & 1.0 & 8.0 & 32.93 & 0.212 \\
\hline
Cu & 8 & 21 & 1.0 & 11.0 & 1.80 & 0.013 \\
Ag & 12 & 24 & 1.0 & 10.0 & 1.27 & 0.008 \\
Au & 6 & 24 & 1.0 & 11.0 & 3.04 & 0.027 \\
\hline
Zn & 8 & 18 & 1.5 & 10.0 & 3.37 & 0.018 \\
Cd & 9 & 15 & 1.0 & 8.0 & 2.29 & 0.013 \\
Hg & 13 & 21 & 0.5 & 12.0 & 1.97 & 0.010 \\
\hline
Al & 7  & 21 & 1.0 & 12.0 & 3.12 & 0.021 \\
Ga & 11 & 18 & 0.5 & 10.0 & 2.27 & 0.021 \\
In & 12 & 15 & 0.5 & 11.0 & 1.89 & 0.018 \\
Tl & 10 & 15 & 0.5 & 12.0 & 3.31 & 0.019 \\
\end{tabular}
\end{ruledtabular}
\end{table}
\endgroup
\begingroup
\squeezetable
\begin{table}[tbp]
\begin{ruledtabular}
\caption{
Optimal cutoff radii, optimal number of Gaussian radial functions and optimal parameters for identifying the set of Gaussian radial functions and the RMSEs of optimized pairwise-MLIP2 for the 31 elemental metals.
}
\label{Table-Optimal2}
\begin{tabular}{ccccccc}
Element & $R_c$ & $N_{\rm rad}$& $p$ & $q_{\rm max}$ & RMSE (energy) & RMSE (force) \\
& (\AA) & & (\AA$^{-2}$) & (\AA) & (meV/atom) & (eV/\AA) \\
\hline
Li & 12 & 24 & 1.5 & 11.0 & 0.16 & 0.002 \\
Na & 12 & 24 & 1.5 & 10.0 & 0.13 & 0.001 \\
K & 13 & 24 & 1.0 & 12.0 & 0.13 & 0.001 \\
Rb & 14 & 24 & 1.5 & 12.0 & 0.18 & 0.001 \\
Cs & 14 & 24 & 0.5 & 12.0 & 0.14 & 0.001 \\
\hline
Be & 7 & 21 & 2.0 & 8.0 & 2.93 & 0.033 \\
Mg & 10 & 21 & 1.5 & 9.0 & 0.48 & 0.005 \\
Ca & 13 & 24 & 2.0 & 10.0 & 0.73 & 0.006 \\
Sr & 12 & 24 & 1.5 & 11.0 & 0.55 & 0.006 \\
Ba & 10 & 21 & 1.0 & 9.0 & 0.93 & 0.010 \\
\hline
Sc & 14 & 24 & 2.0 & 10.0 & 2.55 & 0.036 \\
Y & 11 & 24 & 2.0 & 9.0 & 2.70 & 0.034 \\
\hline
Ti & 13 & 24 & 2.0 & 9.0 & 3.83 & 0.073 \\
Zr & 9 & 21 & 2.0 & 8.0 & 4.71 & 0.071 \\
Hf & 8 & 15 & 1.5 & 7.0 & 6.04 & 0.095 \\
\hline
V & 7 & 24 & 2.0 & 7.0 & 7.72 & 0.093 \\
Nb & 8 & 24 & 2.0 & 7.0 & 7.19 & 0.095 \\
Ta & 7 & 18 & 2.0 & 7.0 & 9.36 & 0.114 \\
\hline
Cr & 7 & 24 & 2.0 & 7.0 & 6.99 & 0.097 \\
Mo & 6 & 21 & 2.0 & 8.0 & 7.88 & 0.125 \\
W & 6 & 18 & 1.5 & 9.0 & 13.36 & 0.158 \\
\hline
Cu & 9 & 24 & 2.0 & 8.0 & 0.70 & 0.008 \\
Ag & 10 & 24 & 2.0 & 11.0 & 0.60 & 0.005 \\
Au & 6 & 12 & 2.0 & 8.0 & 1.78 & 0.022 \\
\hline
Zn & 8 & 18 & 2.0 & 8.0 & 1.03 & 0.010 \\
Cd & 13 & 24 & 1.5 & 10.0 & 0.65 & 0.006 \\
Hg & 10 & 24 & 1.5 & 9.0 & 0.67 & 0.007 \\
\hline
Al & 10 & 24 & 2.0 & 9.0 & 0.89 & 0.011 \\
Ga & 11 & 24 & 2.0 & 8.0 & 0.80 & 0.012 \\
In & 12 & 24 & 1.0 & 10.0 & 0.81 & 0.011 \\
Tl & 10 & 24 & 2.0 & 9.0 & 0.78 & 0.010 \\
\end{tabular}
\end{ruledtabular}
\end{table}
\endgroup
\begingroup
\squeezetable
\begin{table}[tbp]
\begin{ruledtabular}
\caption{
Optimal cutoff radii, optimal number of Gaussian radial functions, and optimal parameters for identifying the set of Gaussian radial functions and the RMSEs of the optimized angular-dependent MLIP for the 31 elemental metals.
}
\label{Table-Optimal3}
\begin{tabular}{ccccccc}
Element & $R_c$ & $N_{\rm rad}$& $p$ & $q_{\rm max}$ & RMSE (energy) & RMSE (force) \\
& (\AA) & & (\AA$^{-2}$) & (\AA) & (meV/atom) & (eV/\AA) \\
\hline
Li & 12 & 24 & 1.0 & 11.0 & 0.08 & 0.001 \\
Na & 12 & 24 & 1.0 & 12.0 & 0.07 & 0.000 \\
K & 13 & 24 & 0.5 & 10.0 & 0.11 & 0.000 \\
Rb & 14 & 24 & 1.0 & 12.0 & 0.15 & 0.000 \\
Cs & 14 & 24 & 0.5 & 9.0 & 0.10 & 0.000 \\
\hline
Be & 7 & 24 & 1.5 & 7.0 & 0.65 & 0.009 \\
Mg & 10 & 24 & 1.5 & 8.0 & 0.33 & 0.001 \\
Ca & 13 & 24 & 1.5 & 7.0 & 0.36 & 0.001 \\
Sr & 12 & 24 & 0.5 & 10.0 & 0.10 & 0.001 \\
Ba & 10 & 24 & 0.5 & 9.0 & 0.10 & 0.002 \\
\hline
Sc & 14 & 24 & 1.0 & 9.0 & 0.29 & 0.009 \\
Y & 11 & 24 & 1.0 & 8.0 & 0.33 & 0.009 \\
\hline
Ti & 13 & 24 & 1.5 & 8.0 & 0.49 & 0.019 \\
Zr & 9 & 24 & 1.0 & 7.0 & 0.55 & 0.022 \\
Hf & 8 & 24 & 1.5 & 7.0 & 0.67 & 0.023 \\
\hline
V & 7 & 24 & 1.0 & 7.0 & 0.92 & 0.027 \\
Nb & 8 & 24 & 1.5 & 7.0 & 0.80 & 0.029 \\
Ta & 7 & 24 & 0.5 & 7.0 & 1.32 & 0.036 \\
\hline
Cr & 7 & 24 & 1.5 & 8.0 & 1.21 & 0.031 \\
Mo & 6 & 24 & 1.0 & 7.0 & 1.46 & 0.041 \\
W & 6 & 24 & 0.5 & 8.0 & 2.20 & 0.051 \\
\hline
Cu & 9 & 24 & 1.5 & 8.0 & 0.19 & 0.002 \\
Ag & 10 & 24 & 1.0 & 12.0 & 0.29 & 0.001 \\
Au & 6 & 24 & 1.0 & 7.0 & 0.33 & 0.004 \\
\hline
Zn & 8 & 24 & 2.0 & 7.0 & 0.37 & 0.004 \\
Cd & 13 & 24 & 1.5 & 12.0 & 0.27 & 0.002 \\
Hg & 10 & 24 & 1.0 & 10.0 & 0.15 & 0.002 \\
\hline
Al & 10 & 24 & 2.0 & 7.0 & 0.40 & 0.004 \\
Ga & 11 & 24 & 1.0 & 12.0 & 0.39 & 0.004 \\
In & 12 & 24 & 1.0 & 10.0 & 0.46 & 0.004 \\
Tl & 10 & 24 & 2.0 & 8.0 & 0.27 & 0.006 \\
\end{tabular}
\end{ruledtabular}
\end{table}
\endgroup

\section{Phonon dispersion curves predicted by MLIPs}
\begin{figure*}[tbp]
\includegraphics[clip,width=0.85\linewidth]{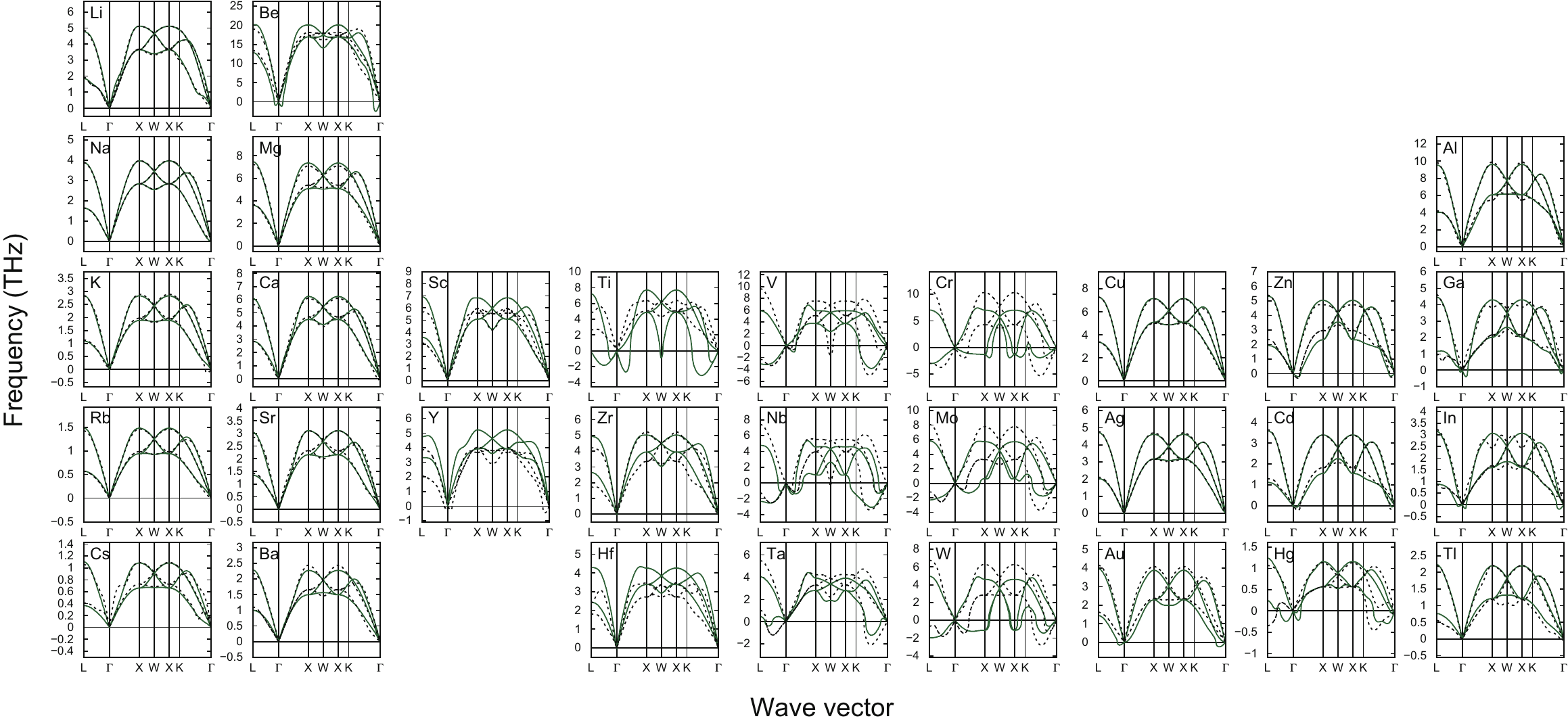}
\includegraphics[clip,width=0.85\linewidth]{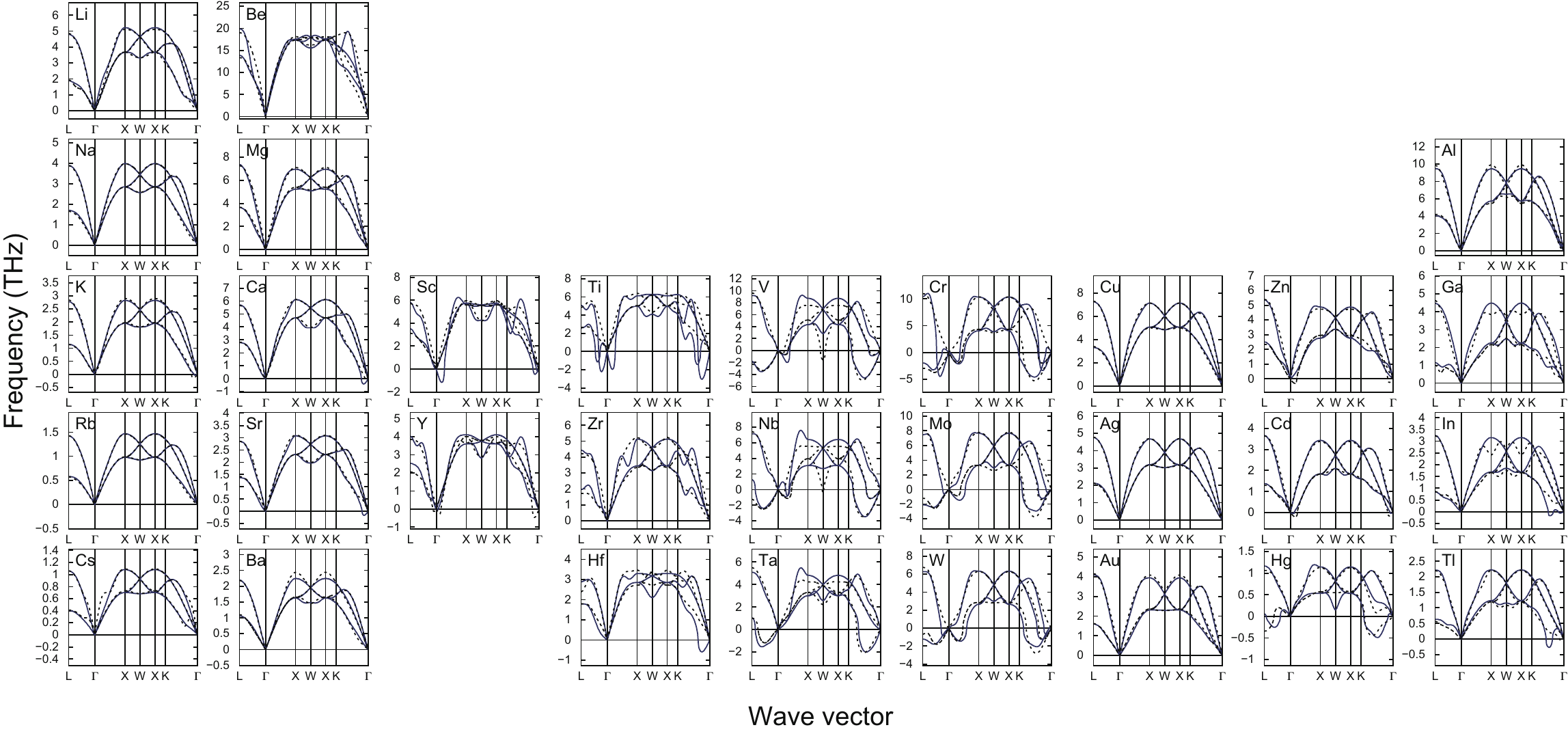}
\includegraphics[clip,width=0.85\linewidth]{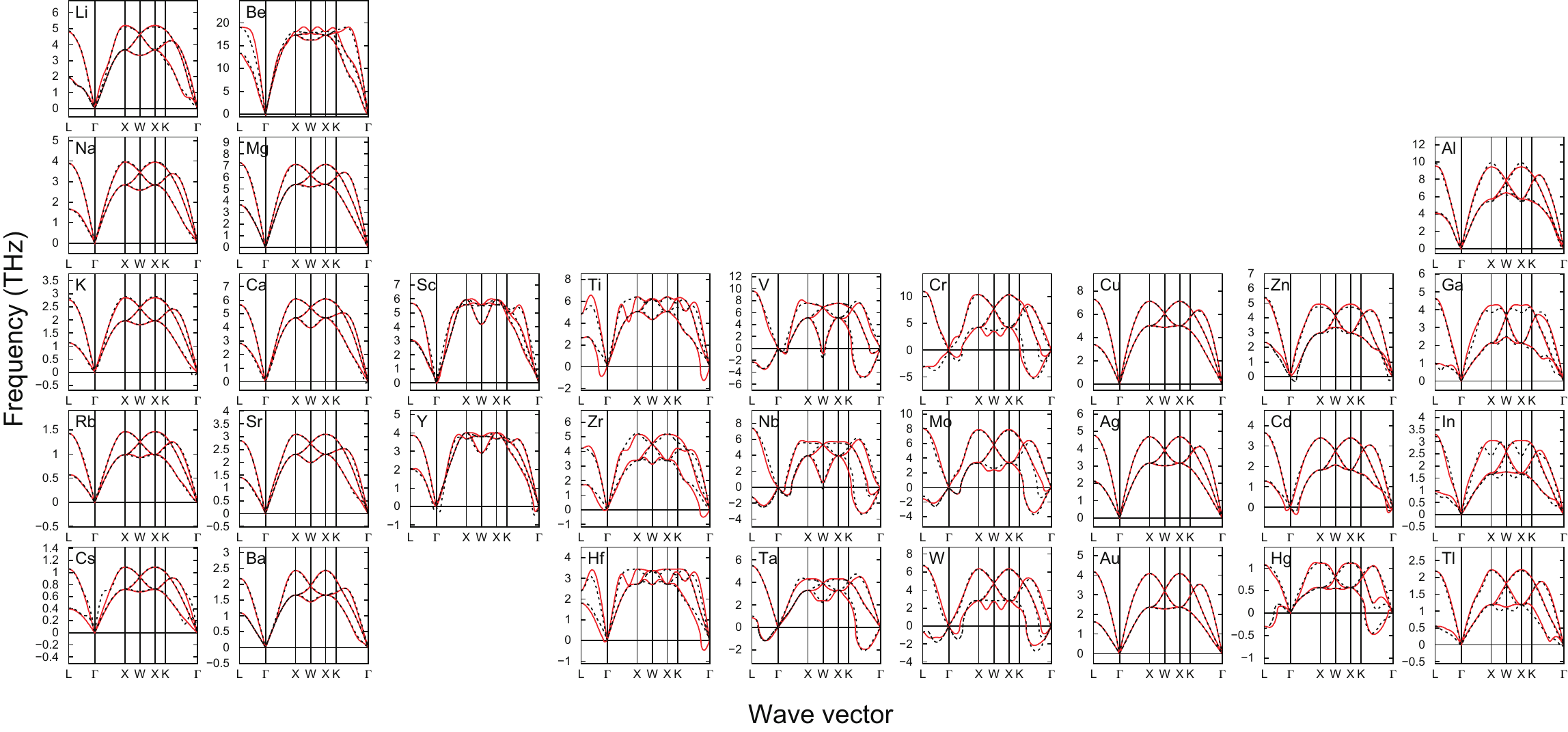}
\caption{ 
Phonon dispersion curves of fcc structure obtained by pairwise-MLIP1 (upper panel), pairwise-MLIP2 (middle panel) and the angular-dependent MLIP (lower panel), shown by the solid lines.
Phonon dispersion curves obtained by the DFT calculation are also shown by the broken lines.
}
\label{Fig-PhononFcc}
\end{figure*}

\begin{figure*}[tbp]
\includegraphics[clip,width=0.85\linewidth]{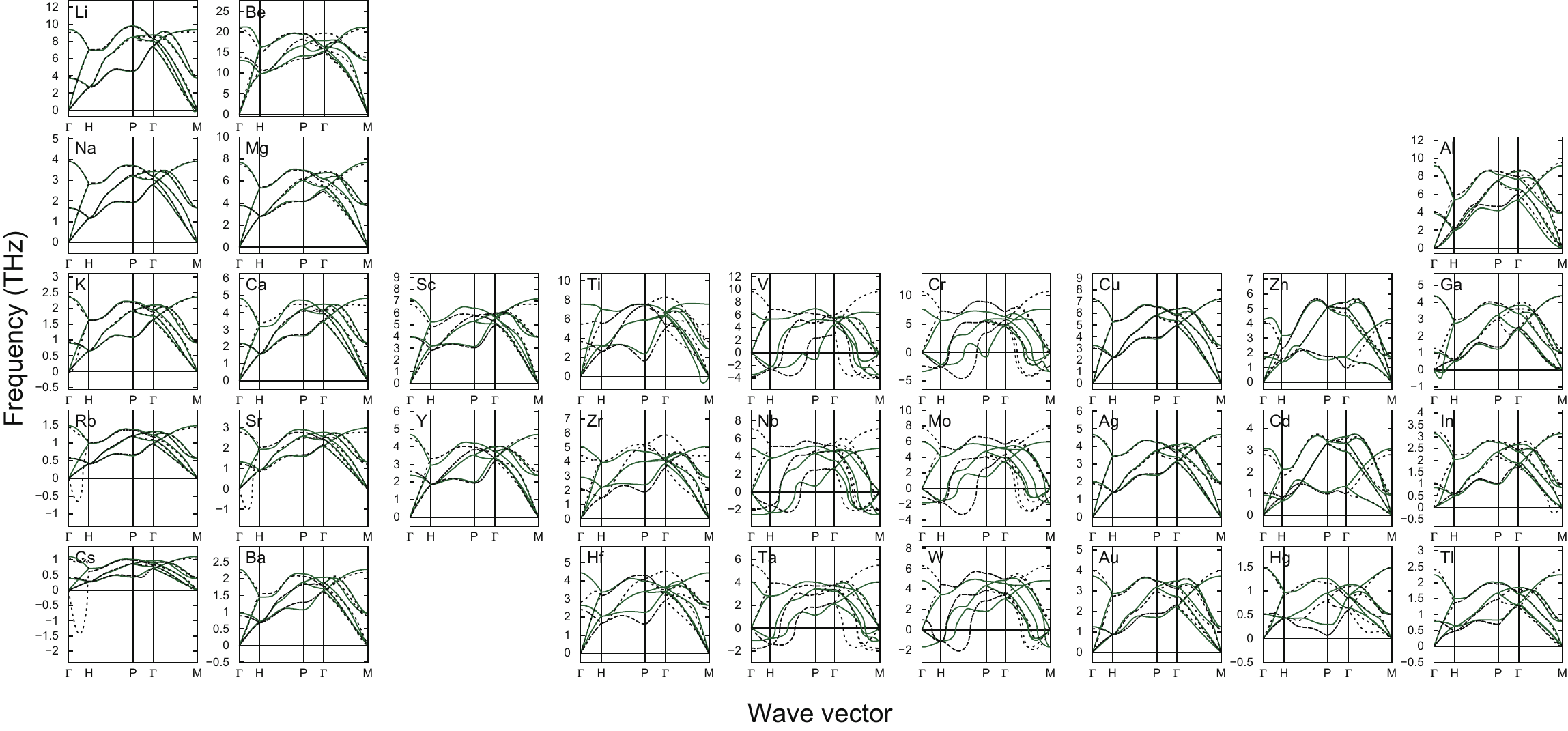}
\includegraphics[clip,width=0.85\linewidth]{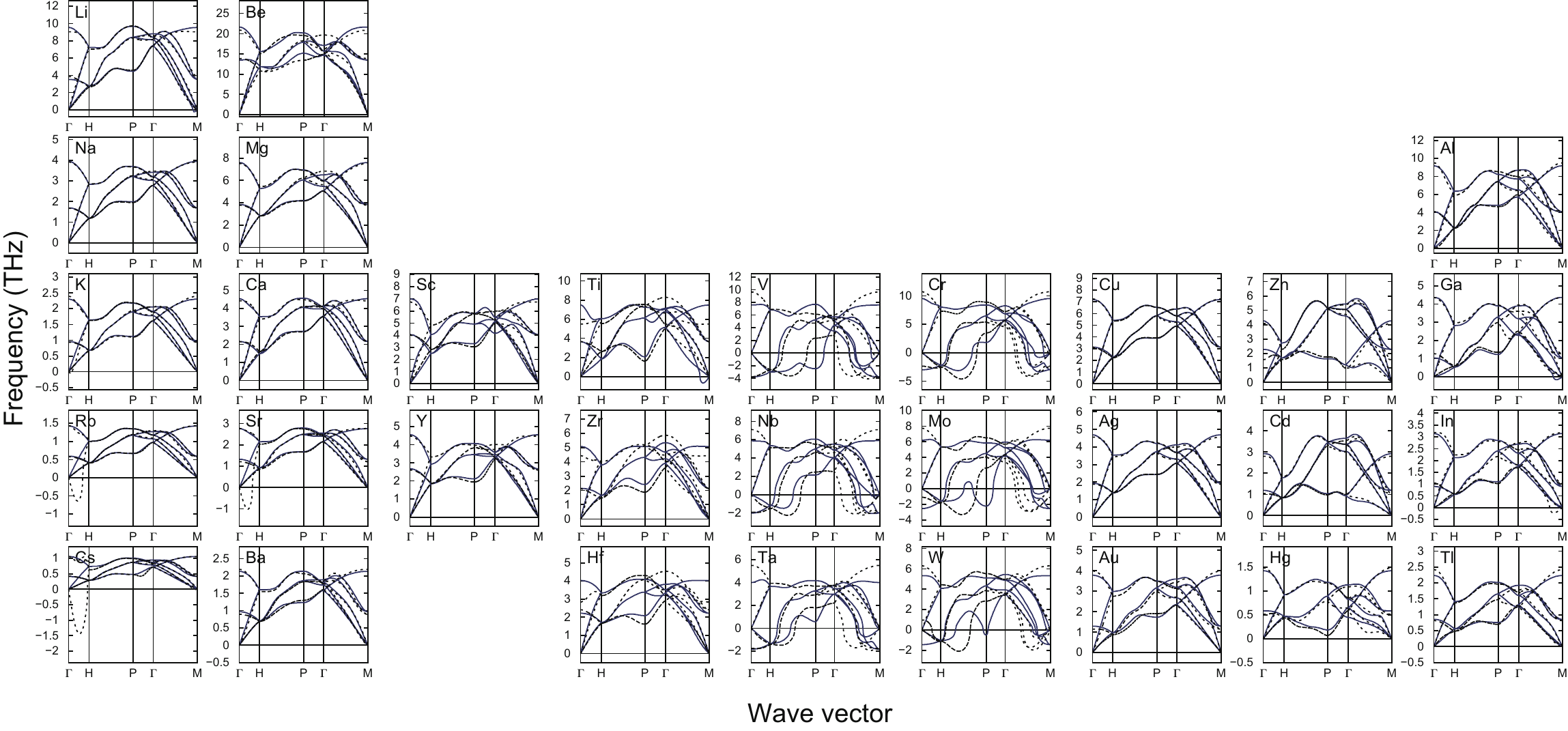}
\includegraphics[clip,width=0.85\linewidth]{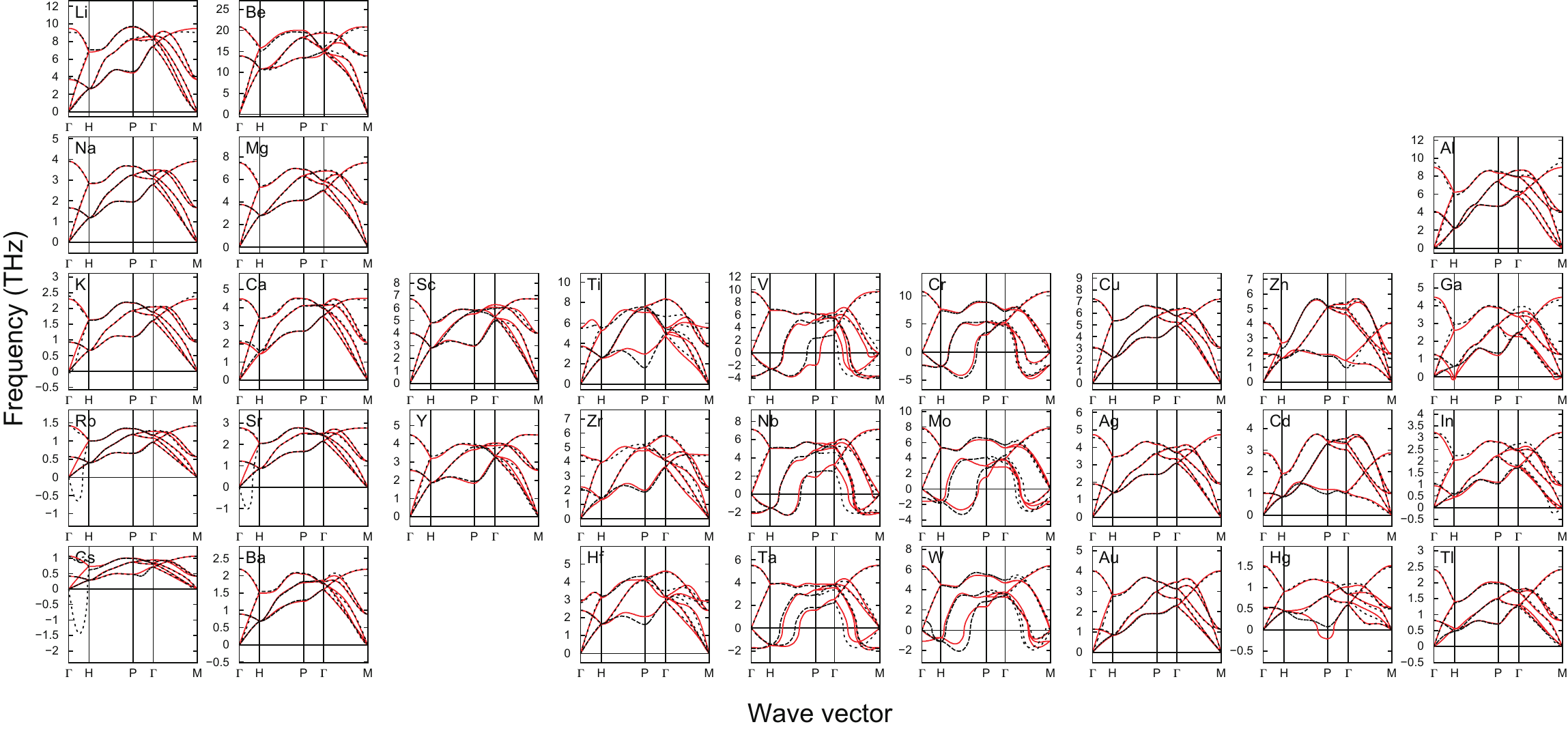}
\caption{ 
Phonon dispersion curves of hcp structure obtained by pairwise-MLIP1 (upper panel), pairwise-MLIP2 (middle panel) and the angular-dependent MLIP (lower panel), shown by the solid lines.
Phonon dispersion curves obtained by the DFT calculation are also shown by the broken lines.
}
\label{Fig-PhononHcp}
\end{figure*}

\begin{figure*}[tbp]
\includegraphics[clip,width=0.85\linewidth]{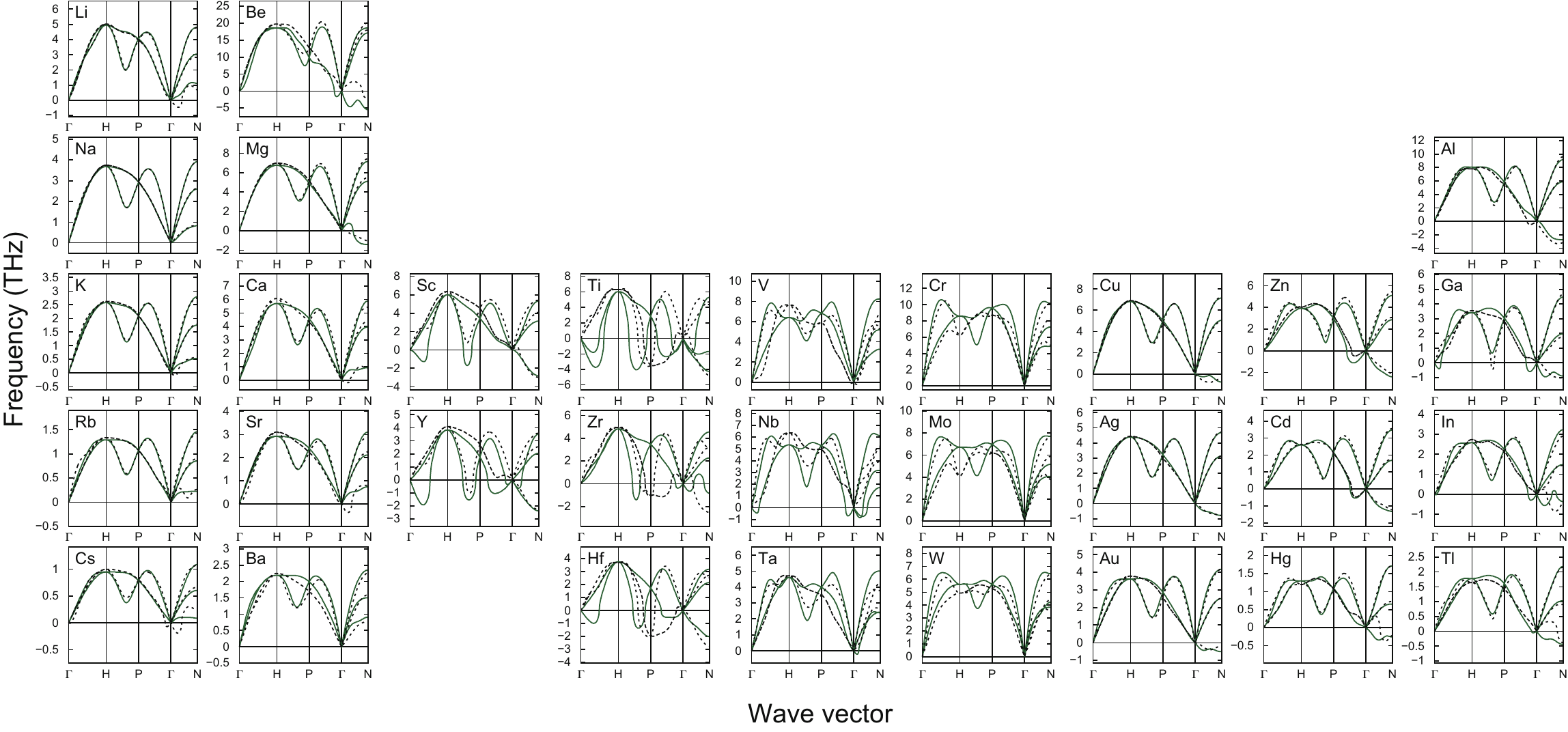}
\includegraphics[clip,width=0.85\linewidth]{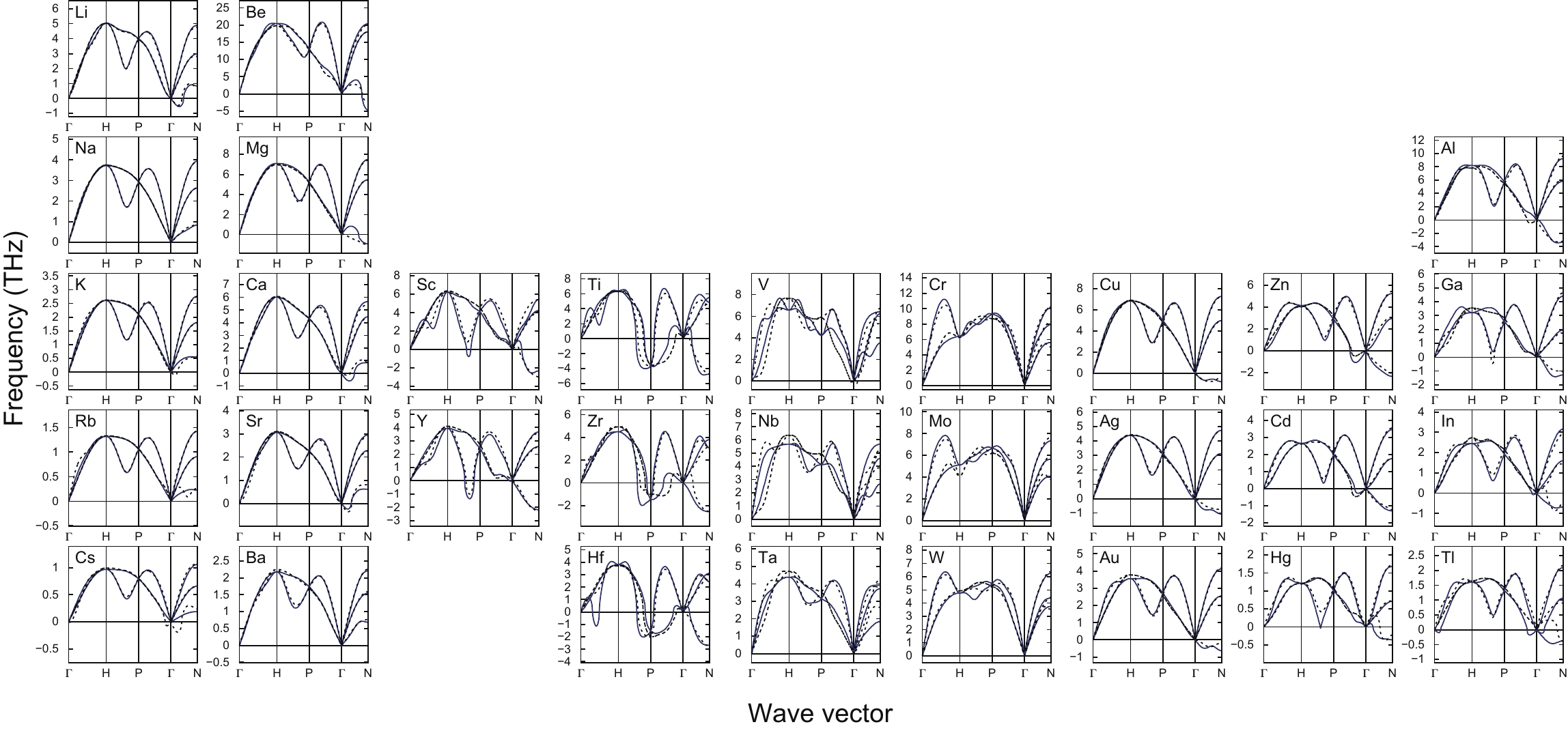}
\includegraphics[clip,width=0.85\linewidth]{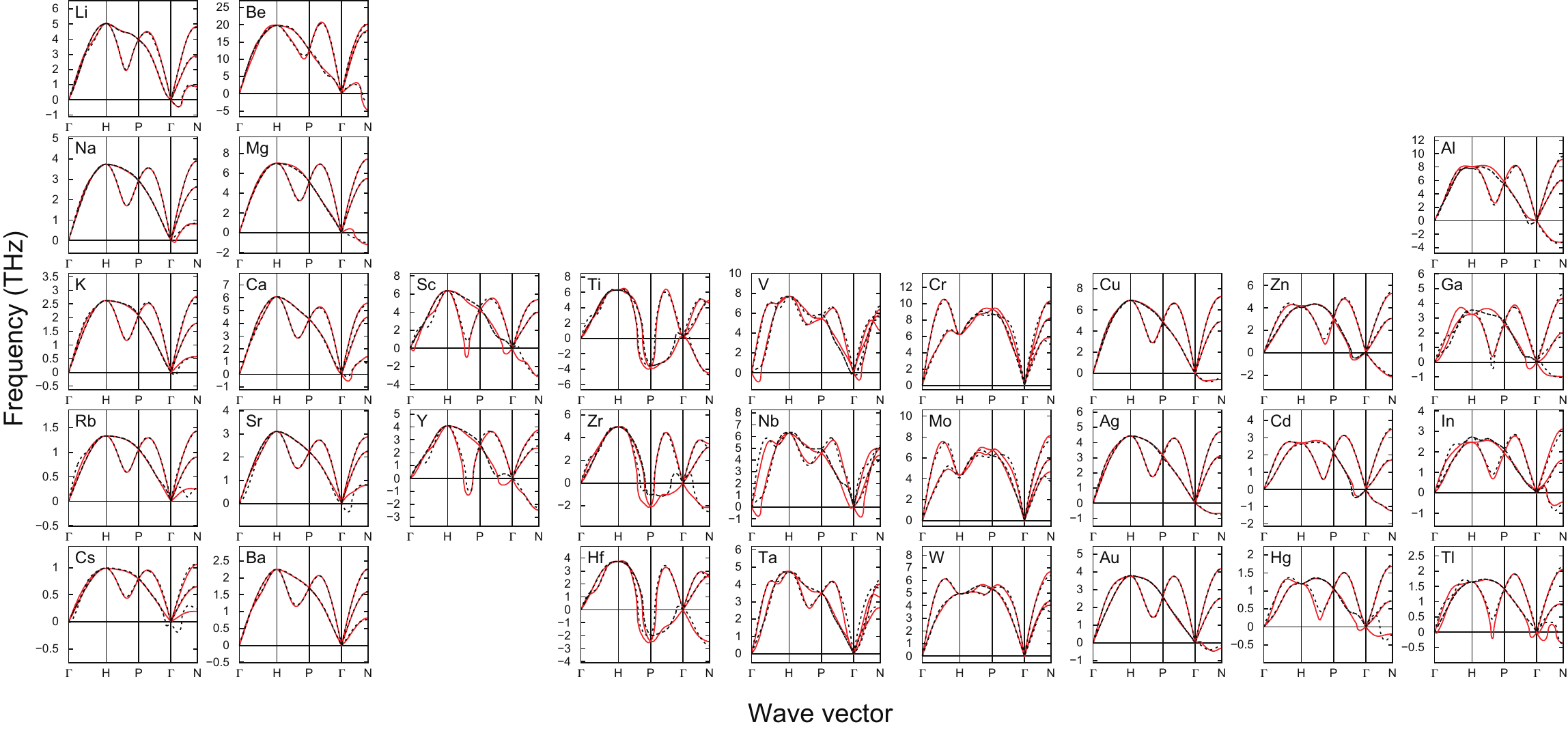}
\caption{ 
Phonon dispersion curves of bcc structure obtained by pairwise-MLIP1 (upper panel), pairwise-MLIP2 (middle panel) and the angular-dependent MLIP (lower panel), shown by the solid lines.
Phonon dispersion curves obtained by the DFT calculation are also shown by the broken lines.
}
\label{Fig-PhononBcc}
\end{figure*}

\begin{figure*}[tbp]
\includegraphics[clip,width=0.85\linewidth]{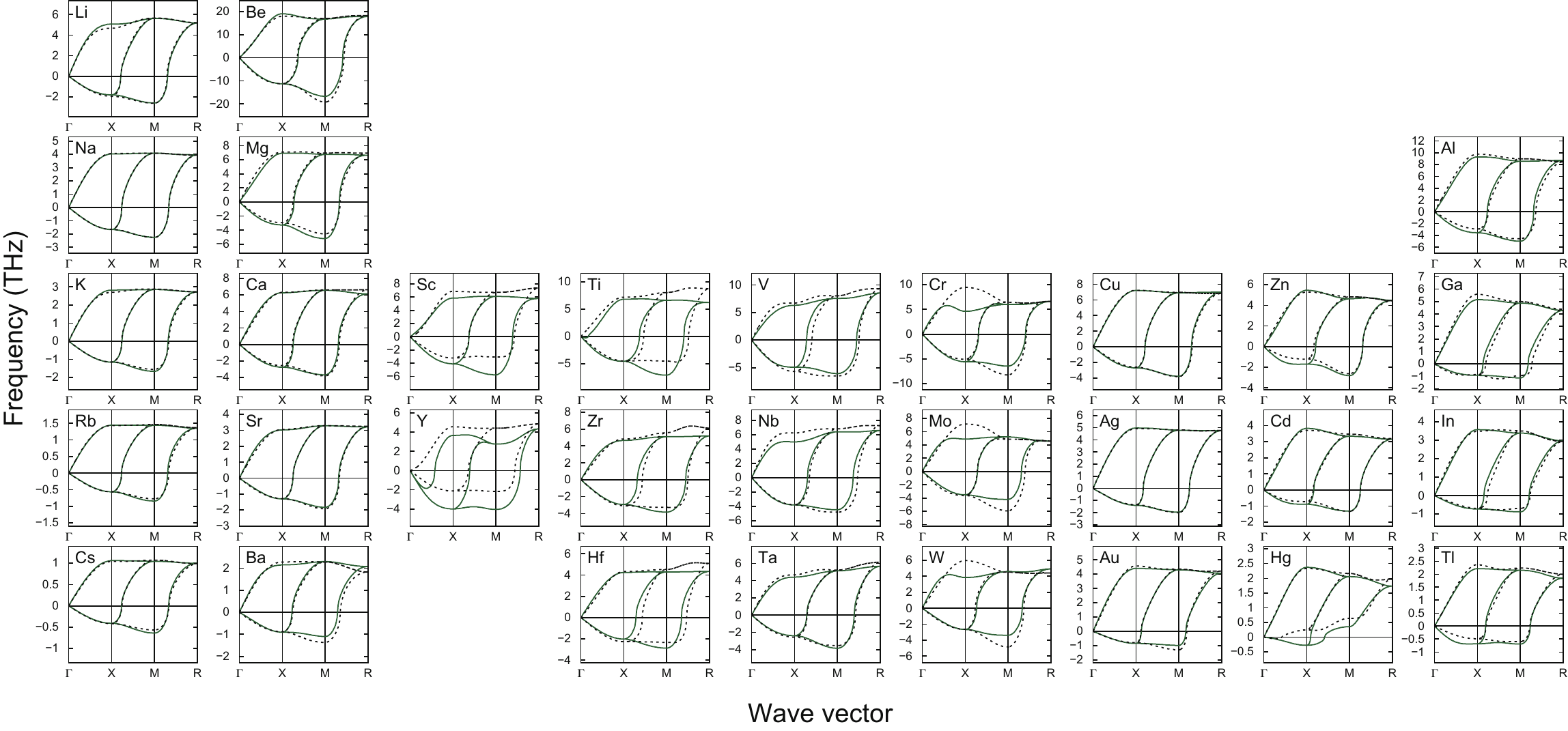}
\includegraphics[clip,width=0.85\linewidth]{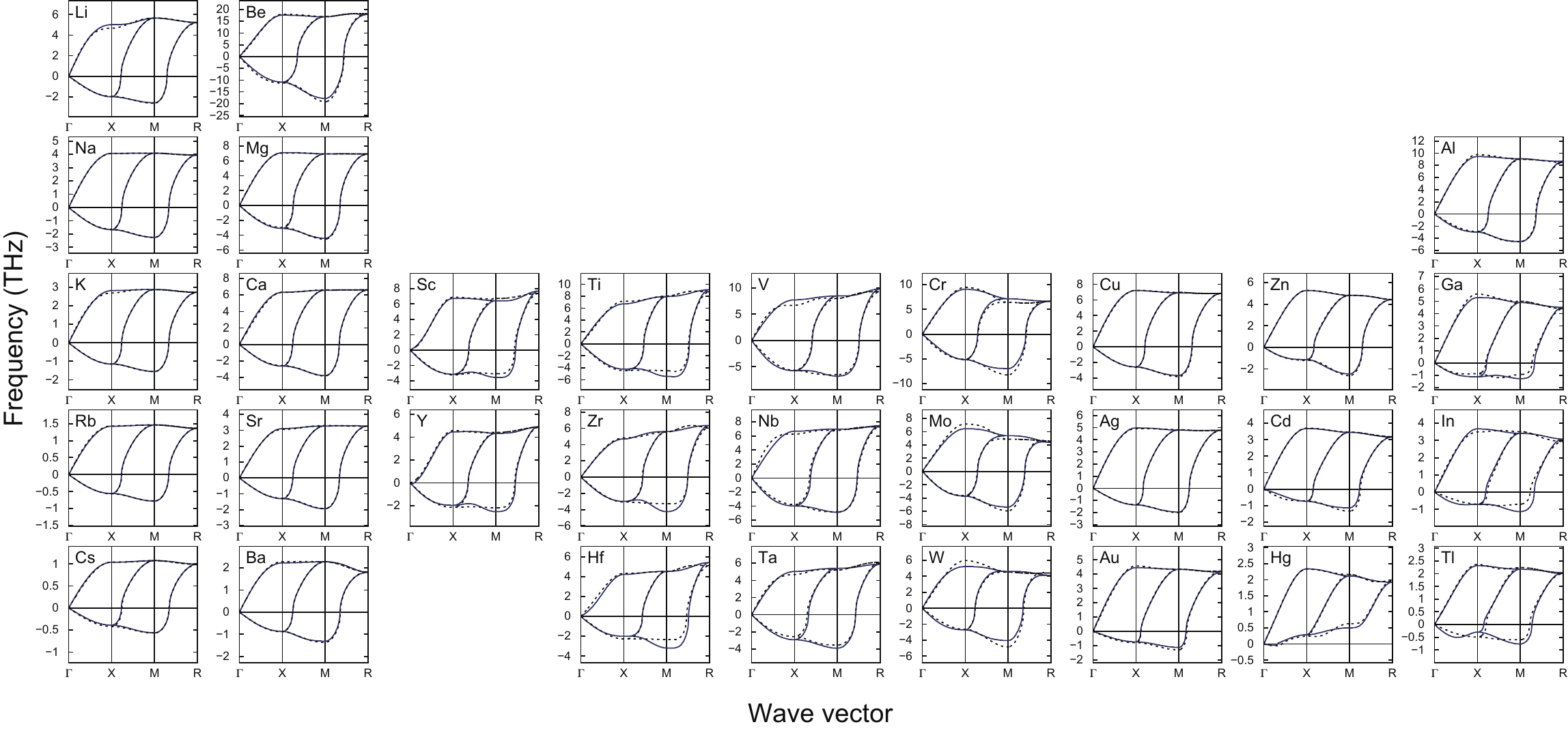}
\includegraphics[clip,width=0.85\linewidth]{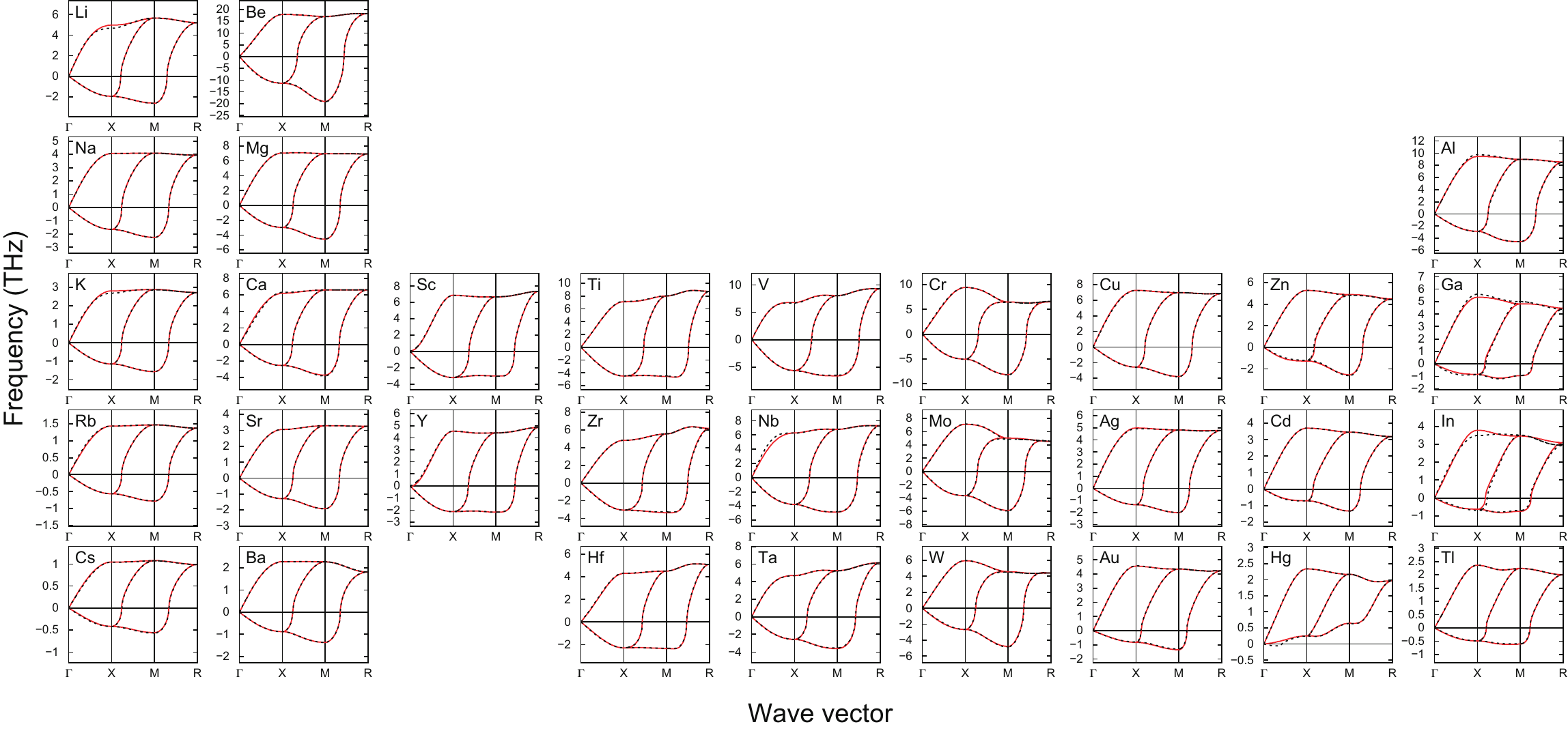}
\caption{ 
Phonon dispersion curves of sc structure obtained by pairwise-MLIP1 (upper panel), pairwise-MLIP2 (middle panel) and the angular-dependent MLIP (lower panel), shown by the solid lines.
Phonon dispersion curves obtained by the DFT calculation are also shown by the broken lines.
}
\label{Fig-PhononSc}
\end{figure*}

\begin{figure*}[tbp]
\includegraphics[clip,width=0.85\linewidth]{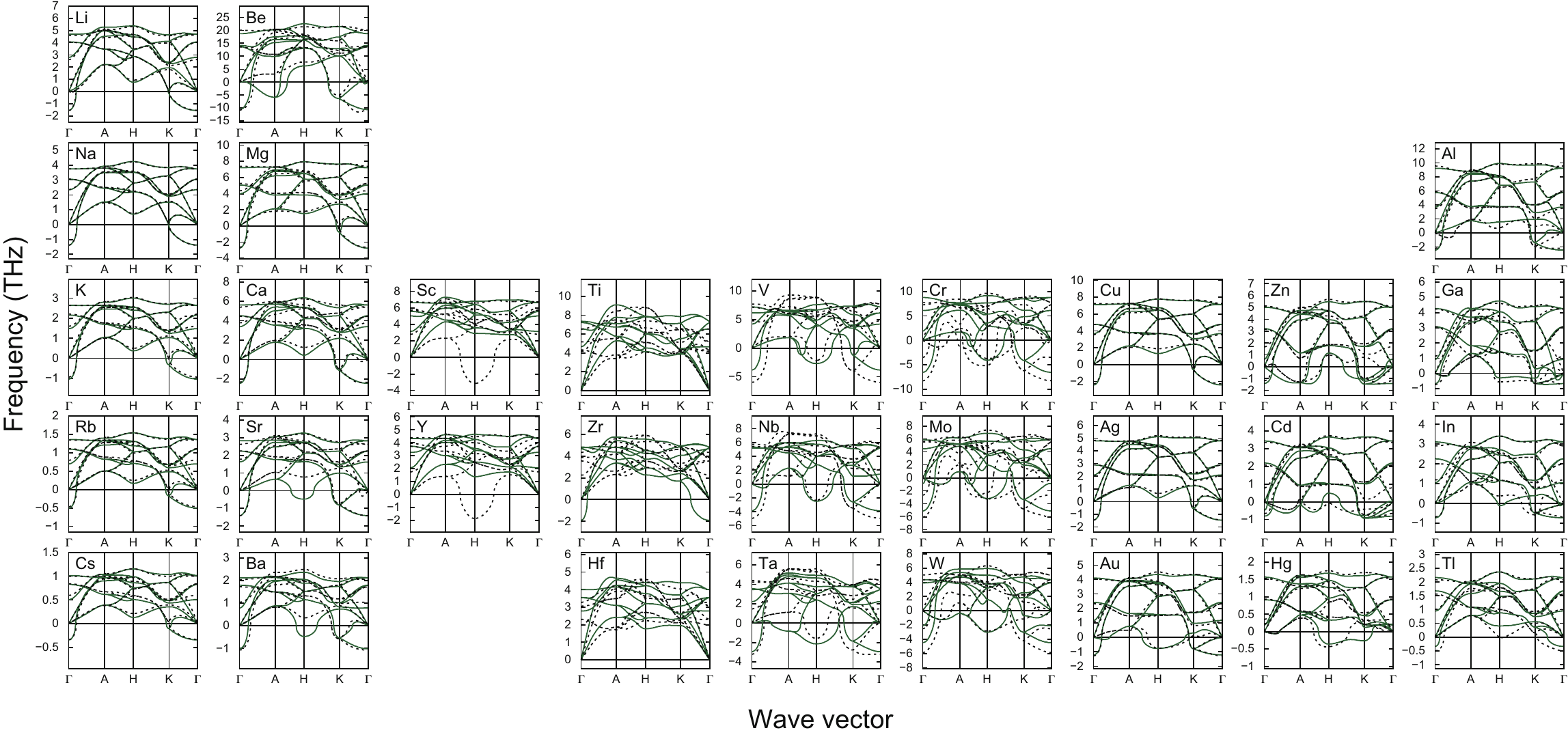}
\includegraphics[clip,width=0.85\linewidth]{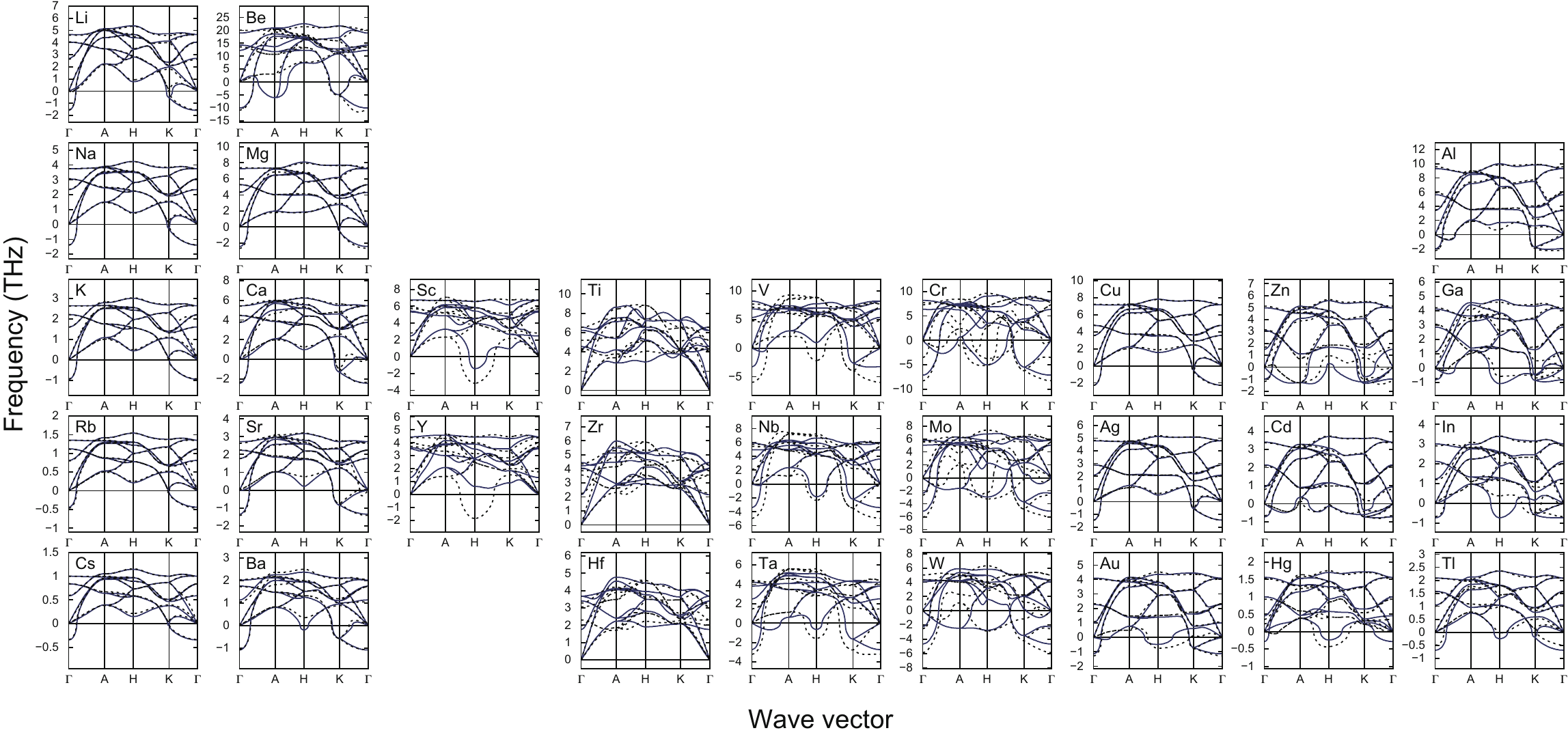}
\includegraphics[clip,width=0.85\linewidth]{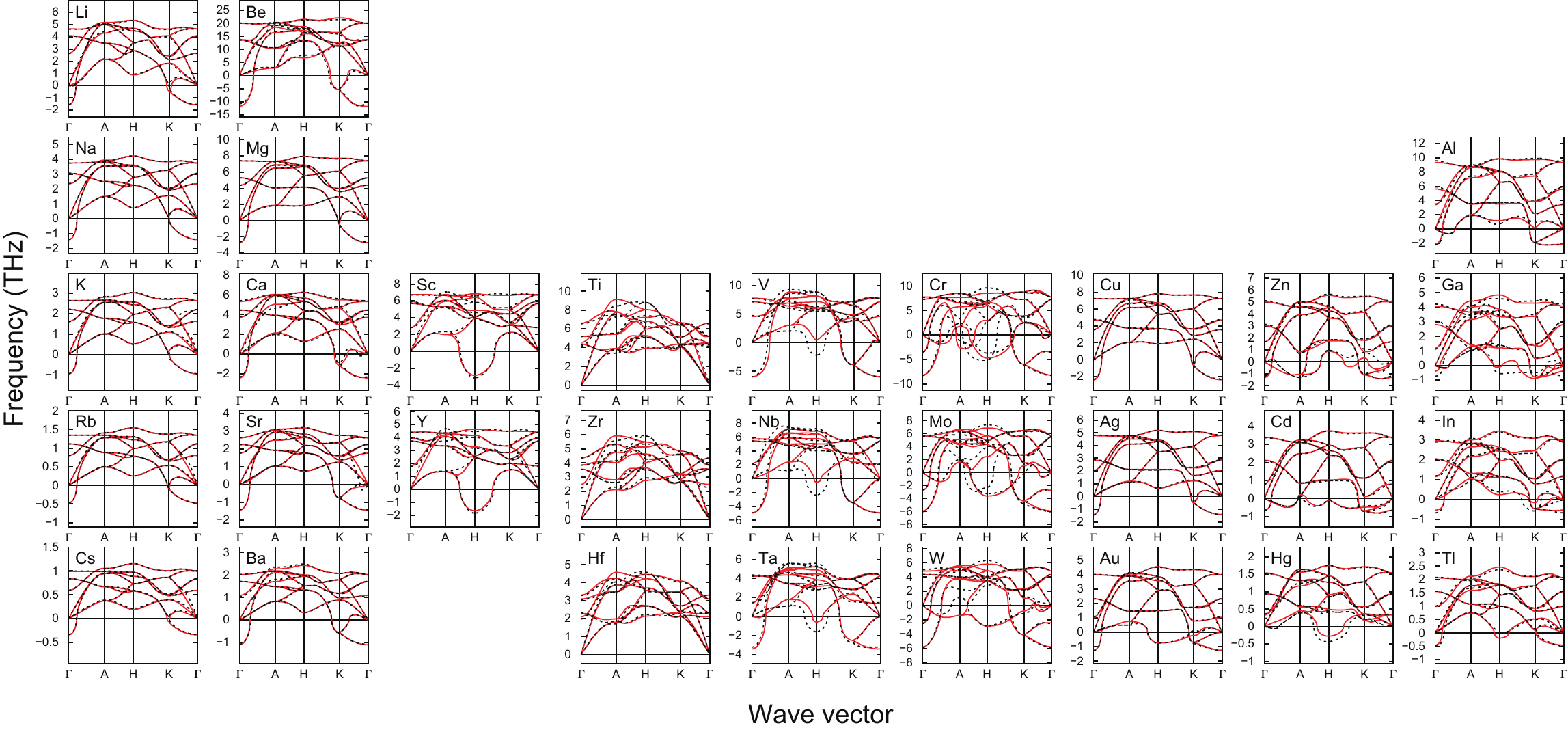}
\caption{ 
Phonon dispersion curves of $\omega$ structure obtained by pairwise-MLIP1 (upper panel), pairwise-MLIP2 (middle panel) and the angular-dependent MLIP (lower panel), shown by the solid lines.
Phonon dispersion curves obtained by the DFT calculation are also shown by the broken lines.
}
\label{Fig-PhononOmega}
\end{figure*}

\begin{figure*}[tbp]
\includegraphics[clip,width=0.85\linewidth]{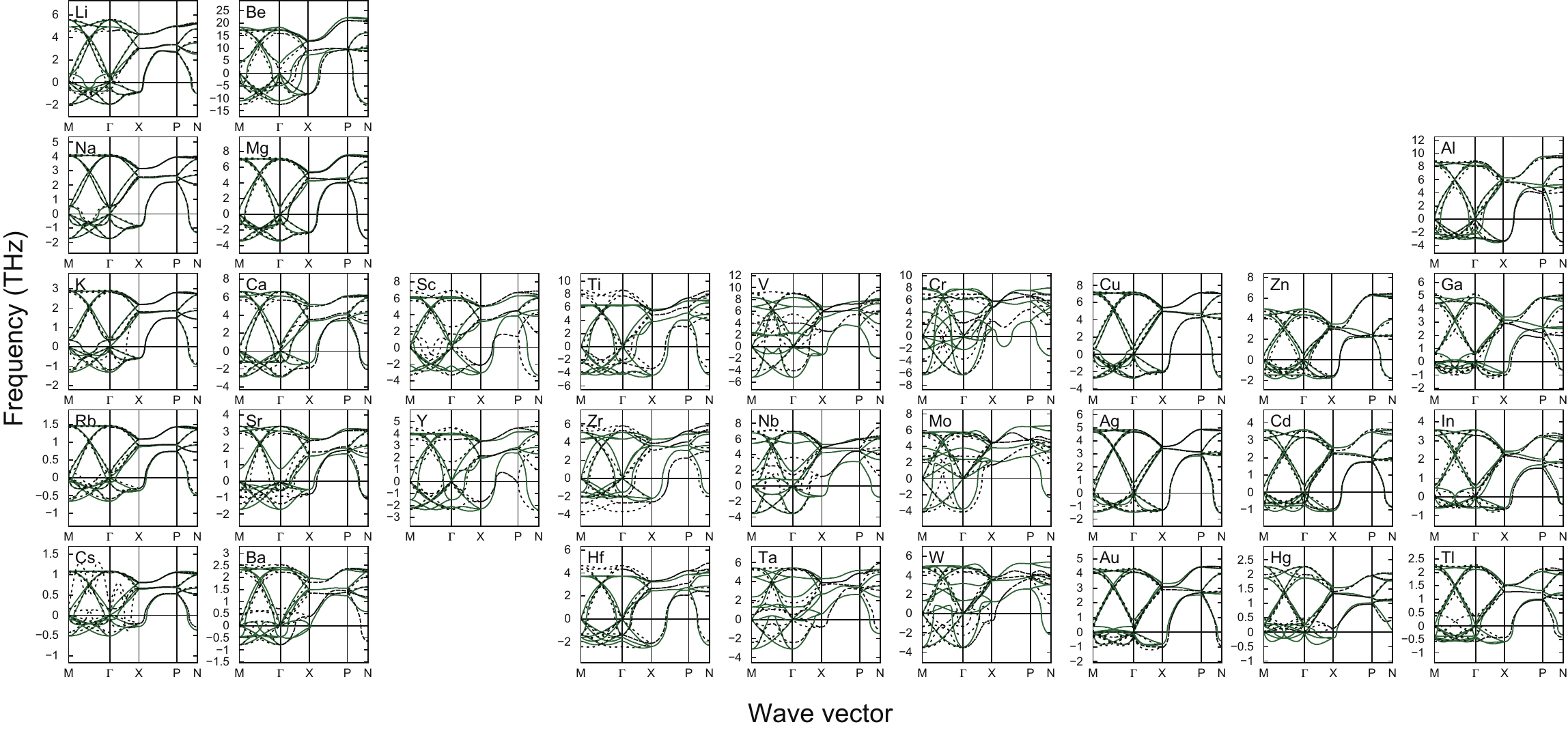}
\includegraphics[clip,width=0.85\linewidth]{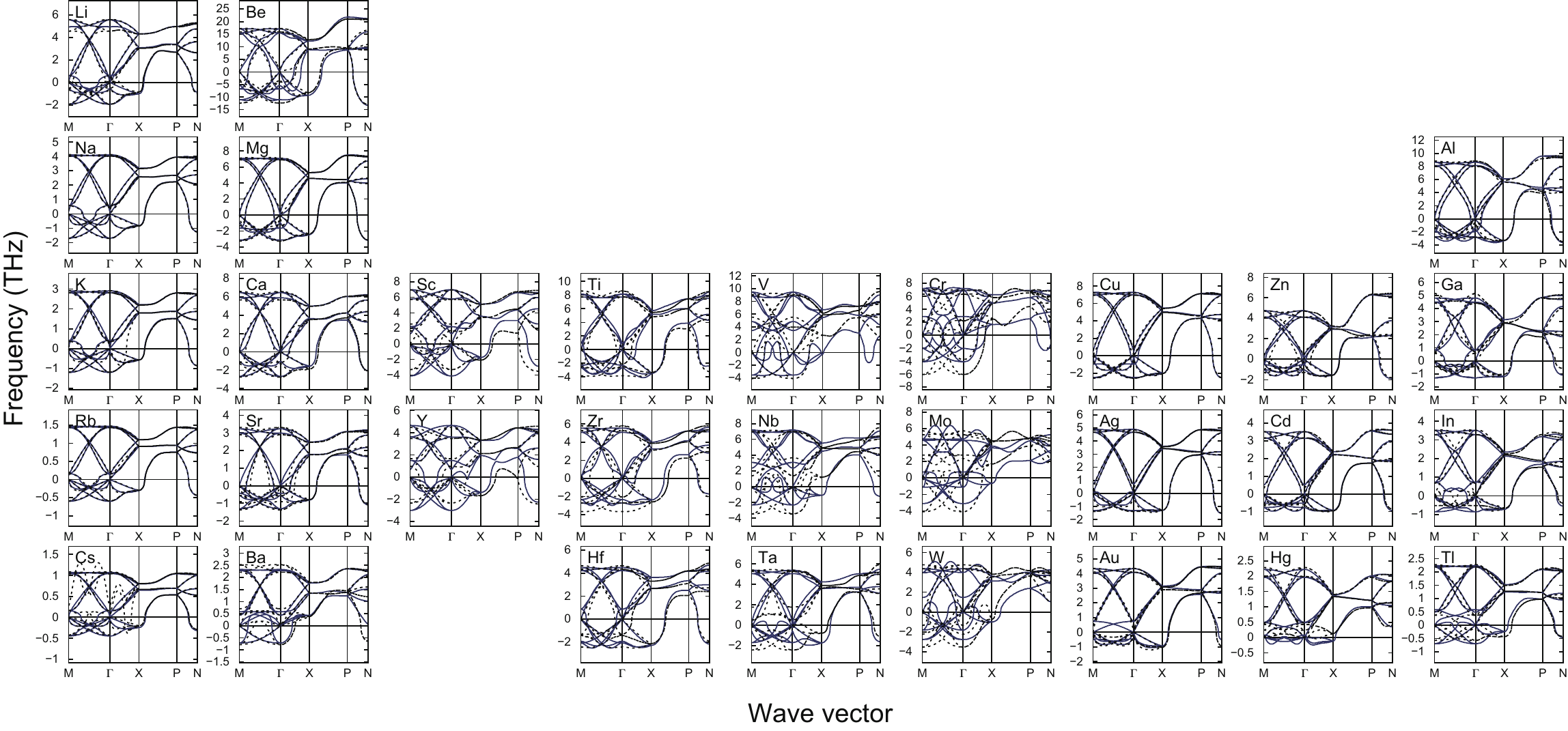}
\includegraphics[clip,width=0.85\linewidth]{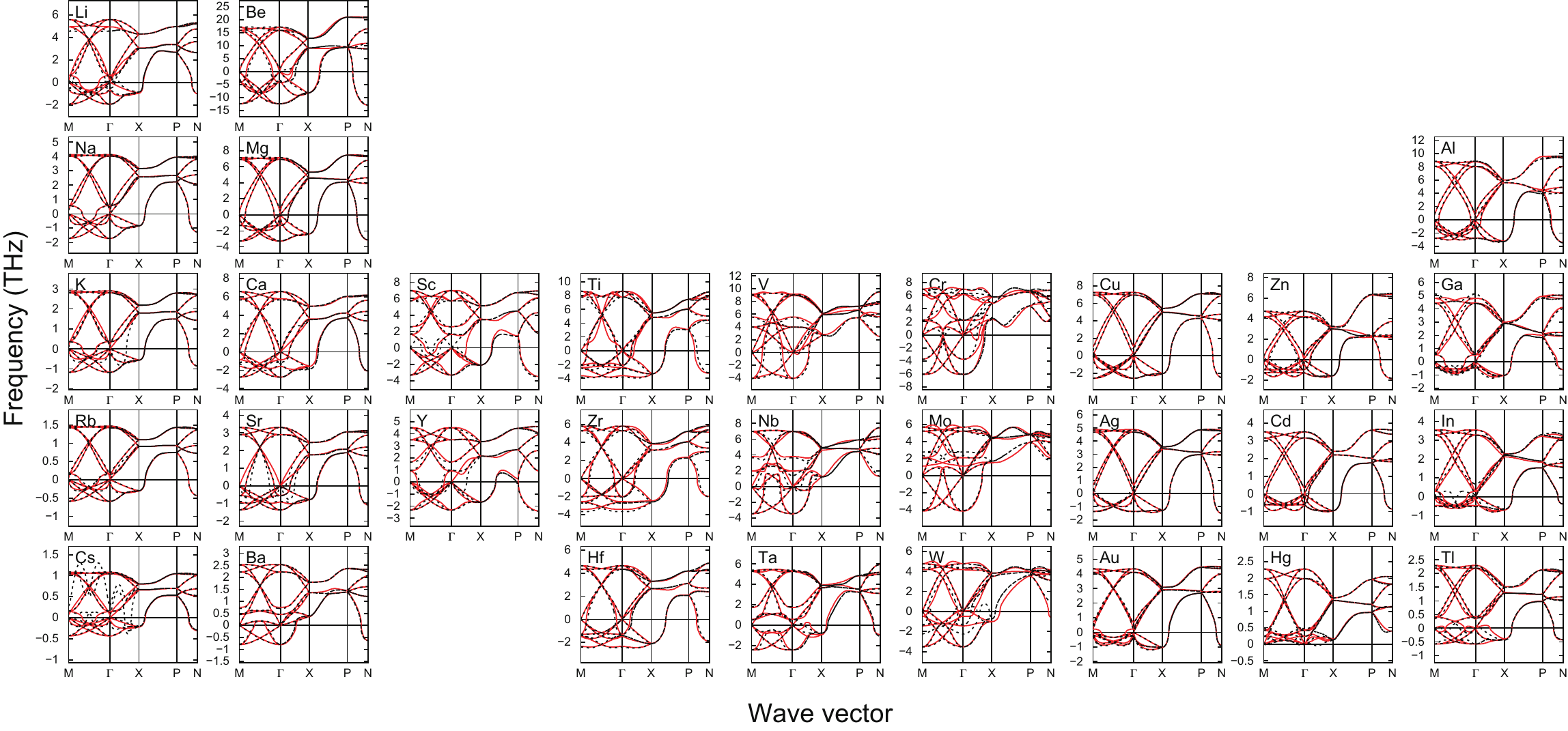}
\caption{ 
Phonon dispersion curves of $\beta$-tin structure obtained by pairwise-MLIP1 (upper panel), pairwise-MLIP2 (middle panel) and the angular-dependent MLIP (lower panel), shown by the solid lines.
Phonon dispersion curves obtained by the DFT calculation are also shown by the broken lines.
}
\label{Fig-PhononBetaTin}
\end{figure*}

Figures \ref{Fig-PhononFcc}--\ref{Fig-PhononBetaTin} show the phonon dispersion curves of fcc, hcp, bcc, sc, $\omega$ and $\beta$-tin structures predicted by the optimal MLIPs for the 31 elemental metals along with those predicted by the DFT calculation.

\clearpage

\section{Elastic constants predicted by MLIPs}
\begin{figure*}[htbp]
\includegraphics[clip,width=0.85\linewidth]{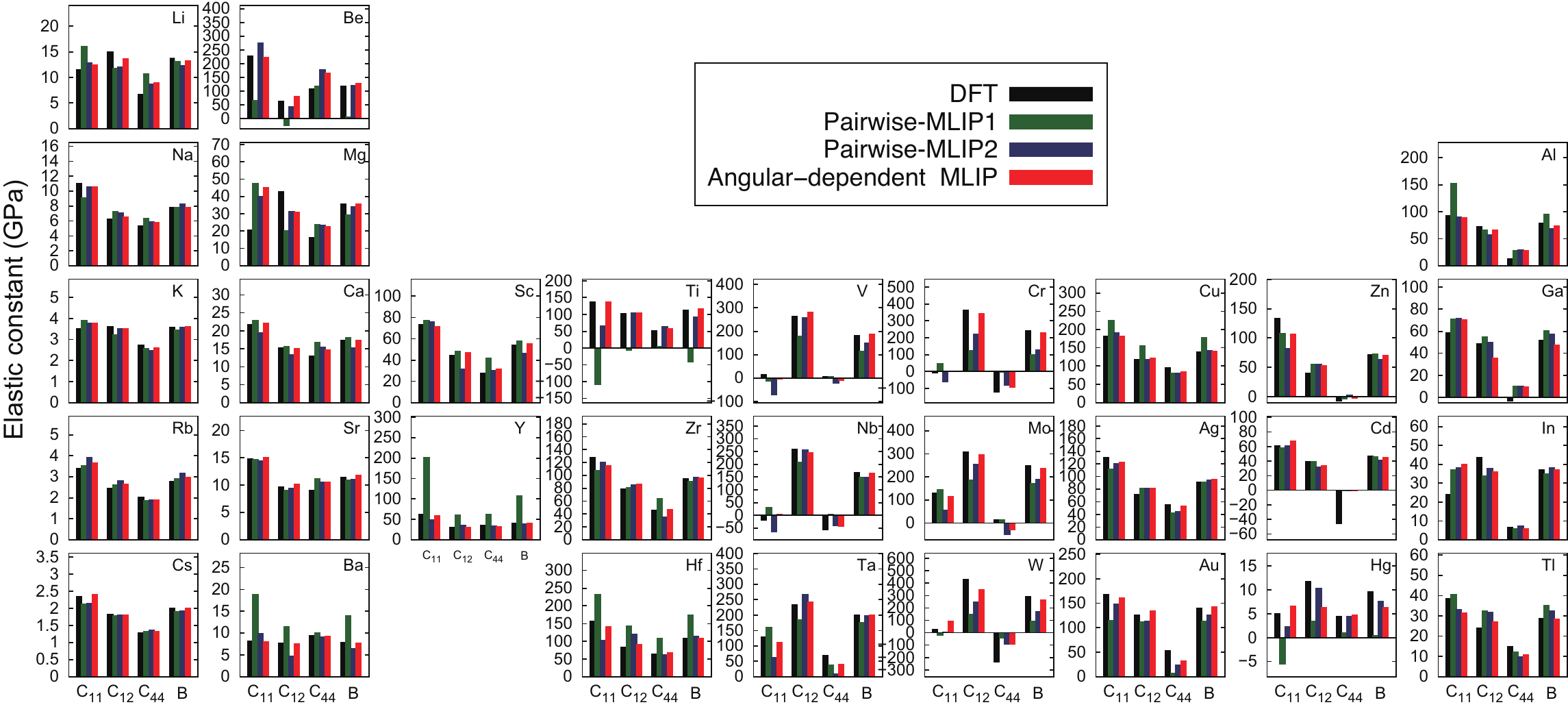}
\includegraphics[clip,width=0.85\linewidth]{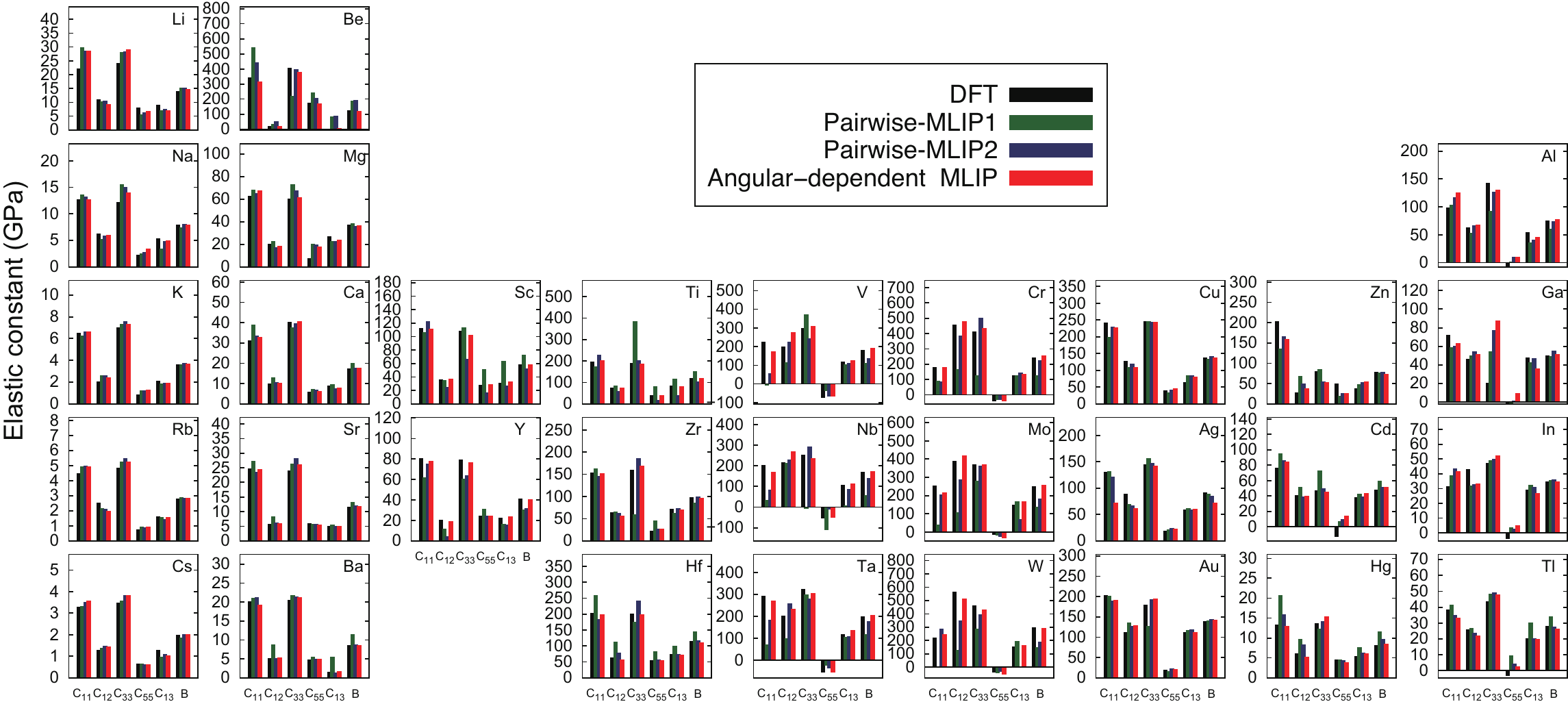}
\includegraphics[clip,width=0.85\linewidth]{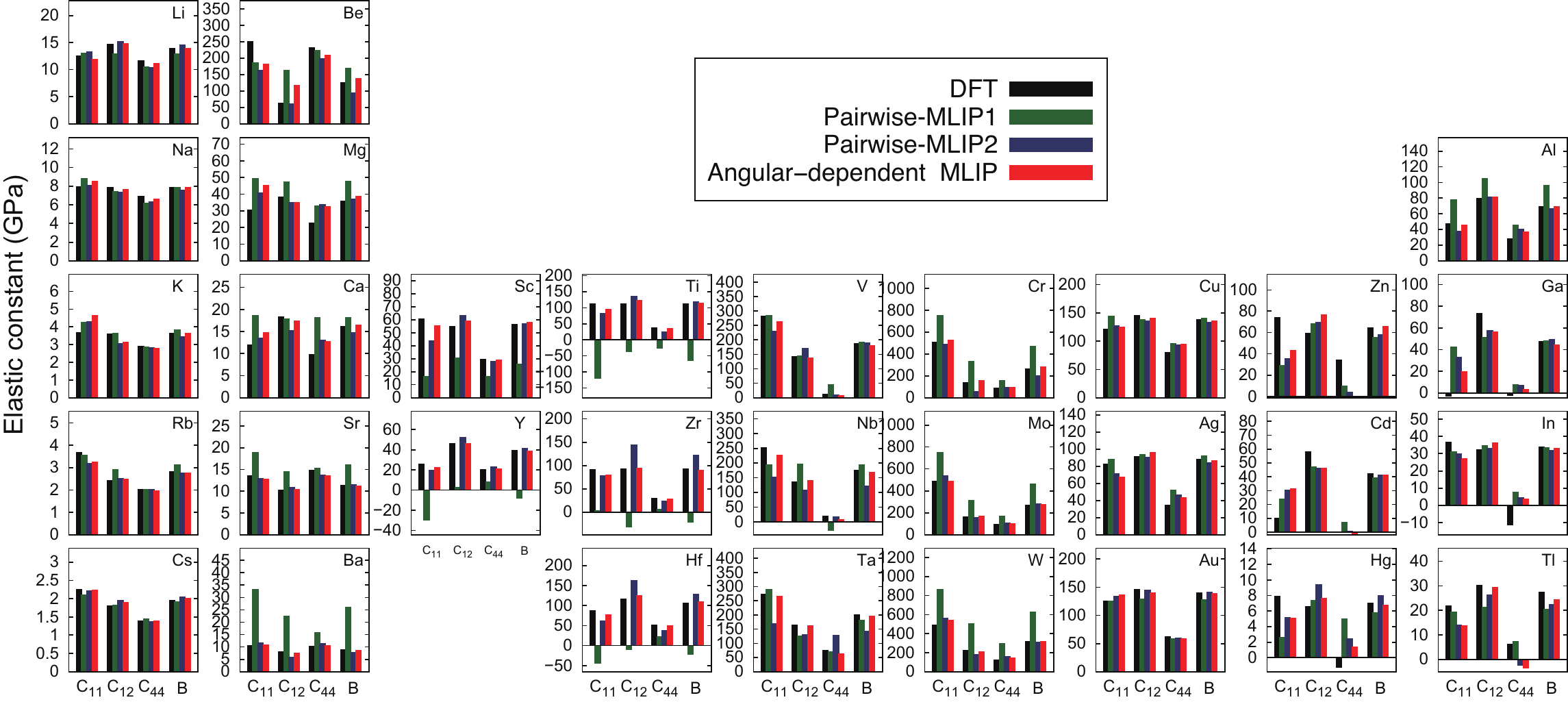}
\caption{ 
Elastic constants and bulk moduli of fcc (upper panel), hcp (middle panel) and bcc (lower panel) structures predicted by MLIPs.
Elastic constants and bulk moduli predicted by the DFT calculation are also shown for comparison.
}
\label{Fig-Elastic1}
\end{figure*}

Figure \ref{Fig-Elastic1} shows the elastic constants and bulk moduli of fcc, hcp, bcc structures predicted by MLIPs for 31 elemental metals along with those predicted by the DFT calculation.
%
%

\bibliography{mlip-31}

\end{document}